\documentclass[aps,pre,twocolumn]{revtex4}
\usepackage{amsmath}
\usepackage{amssymb}
\usepackage{graphicx}
\begin{document}

\title{Diversity and noise effects in a model of homeostatic
  regulation of the sleep-wake cycle}

\author{Marco~Patriarca$^{{\bf 1,2}}$}\email{marcop@ifisc.uib-csic.es}

\author{Svetlana~Postnova$^{{\bf 3,4}}$}\email{postnova@physics.usyd.edu.au}

\author{Hans~A.~Braun$^{{\bf 5}}$}\email{braun@staff.uni-marburg.de}

\author{Emilio~Hern\'andez-Garc\'ia$^{{\bf 1}}$}\email{emilio@ifisc.uib-csic.es}

\author{Ra\'ul~Toral$^{{\bf 1,}}$}\email{raul@ifisc.uib-csic.es}

\affiliation{
{\small{\rm\bf 1} IFISC}, 
Instituto de F\'isica Interdisciplinar y Sistemas Complejos (CSIC-UIB),
Campus Universitat de les Illes Balears, E-07122 Palma de Mallorca, Spain}
\thanks{URL: {\tt http://ifisc.uib-csic.es}}

\affiliation{{\rm\bf 2} National Institute of Chemical Physics and Biophysics,
R\"avala 10, 10143 Tallinn, Estonia}

\affiliation{{\rm\bf 3} School\,of\,Physics, 
The\,University\,of\,Sydney,
Physics\,Annex, A29, 
NSW\,2006\,Sydney, Australia}

\affiliation{{\rm\bf 4} Center for Integrated Research and Understanding of Sleep, 
The University of Sydney, 
Glebe Point Rd, 431, 
NSW 2037 Sydney, Australia}

\affiliation{{\rm\bf 5} Neurodynamics Group, 
Physiology Institute, Marburg University, 
Deutschhausstr. 2, 
D-35037 Marburg, Germany}

\begin{abstract}
\begin{center}
{\bf Abstract}
\end{center}

Recent advances in sleep neurobiology have allowed development 
of physiologically based mathematical models of sleep regulation 
that account for the neuronal dynamics responsible for the regulation 
of sleep-wake cycles and allow 
detailed examination of the underlying mechanisms. Neuronal systems 
in general, and those involved in sleep regulation in particular, 
are noisy and heterogeneous by their nature. It has been shown in 
various systems that certain levels of noise and diversity can 
significantly improve signal encoding. However, these phenomena, 
especially the effects of diversity, are rarely considered in the 
models of sleep regulation. The present paper is focused on a 
neuron-based physiologically motivated model of sleep-wake cycles 
that proposes a novel mechanism of the homeostatic regulation of 
sleep based on the dynamics of a wake-promoting neuropeptide orexin. 
Here this model is generalized by the introduction of intrinsic 
diversity and noise in the orexin-producing neurons, in order 
to study the effect of their presence on the sleep-wake cycle. 
A simple quantitative measure of the quality of a sleep-wake 
cycle is introduced and used to systematically study the generalized 
model for different levels of noise and diversity. The model is 
shown to exhibit a clear diversity-induced resonance:  that is, 
the best wake-sleep cycle turns out to correspond to an intermediate 
level of diversity at the synapses of the orexin-producing neurons. 
On the other hand, only a mild evidence of stochastic resonance is 
found, when the level of noise is varied. These results show that 
disorder, especially in the form of quenched diversity, can be a 
key-element for an efficient or optimal functioning of the homeostatic 
regulation of the sleep-wake cycle. Furthermore, this study provides an 
example of a constructive role of diversity in a neuronal system that 
can be extended beyond the system studied here.
\end{abstract}

\maketitle

\fbox{
\begin{minipage}{3in}
\hfill

\vspace{0.05in}

This paper has been published as:\\
\,\,\\
M. Patriarca, S. Postnova, H.A. Braun, \\
E. Hern\'andez-Garc\'ia, R. Toral,\\
{\it Diversity and Noise Effects in a Model of}\\
{\it Homeostatic Regulation of the Sleep-Wake Cycle},\\ 
PLoS Comput. Biol. {\bf 8}(8): e1002650 (2012)\\
{\tt doi:10.1371/journal.pcbi.1002650}

\vspace{0.05in}

\end{minipage}
}

\pagebreak

\section*{Author Summary}
{\it All biological systems are inherently noisy and heterogeneous. 
Disorder is mostly expected to disturb proper functioning of a 
system, like it can be the case with noise in a radio signal. 
However, it has been demonstrated by numerous studies that 
noise can actually improve signal encoding --– the so-called 
stochastic resonance phenomenon. Recently, it was discovered that 
quenched diversity (heterogeneity) can also enhance the response of 
a system to an external perturbation (diversity-induced resonance). 
In this study we investigate the role of noise and diversity in a 
neuronal model of sleep-wake cycles based on the dynamics of 
the wake-promoting orexin neurons that is crucial for stability of 
wake and sleep states. We demonstrate that suitable levels of 
diversity introduced in the orexin neurons can significantly 
improve the quality of the sleep-wake cycle, and may be essential 
for proper sleep-wake periodicity. Noise, on the other hand, 
provides only a mild improvement.}

\newpage
\section{Introduction}

Disorder, which originates from both noise and diversity, 
is naturally present in all biological systems. In neuronal 
systems some examples are the random opening and closing of ion channels, 
the multitude of stochastic input currents in the neurons, and the diversity of shapes, 
sizes, and electrophysiological properties of the neurons 
\cite{vonEconomo2009a,Gerashchenko2004a}. Disorder 
is often considered to be harmful to the systems' functioning and to information 
encoding. However, it was likewise repeatedly demonstrated that a certain 
level of disorder can facilitate signal encoding by enhancing system's response 
to an external stimuli. For instance, quenched diversity clearly shows its 
constructive role in the phenomenon of diversity-induced resonance, in 
which an assembly of heterogeneous excitable units presents an optimal 
response to an external forcing for a suitable intermediate degree of 
heterogeneity \cite{Tessone2006a,Tessone2007a,Komin2010a}. 
Similar constructive effects can be observed 
in the presence of noise. For example, interplay of noise and nonlinear 
forces produces the directed motion of motor proteins  \cite{Astumian1997a}, 
order-disorder transitions, oscillations, and synchronization in assemblies of excitable 
units \cite{Gassmann1997a,Zaks2005a,Lindner2004a}, and an optimized system response 
in the ubiquitous phenomenon of stochastic resonance \cite{Moss1995a,Gammaitoni1998a}, e.g. 
in ion-channels and neurons \cite{Longtin1993a,Braun1994a,Bezrukov1998a,Longtin1997a,Chialvo1997a,Neiman1998a}.

In the present study we examine the effects of noise and diversity 
(heterogeneity) in a physiologically based neuronal model of 
sleep-wake cycles \cite{Postnova2009a}. This model introduces a novel mechanism of 
the homeostatic regulation of sleep based on the dynamics of a 
wake-promoting neuropeptide orexin (also called hypocretin), 
assuming depression of orexinergic synapses during wakefulness 
and their recovery during sleep. This mechanism is based on 
the experimental findings of the essential role of orexin system in 
maintaining wakefulness and its ability to integrate the sleep-wake relevant
information coming from many brain areas \cite{Peyron1998a,Yoshida2006a} and 
respond to changes in the body external and internal environments by encoding the body activity
state, energy balance, sensory and emotional stimuli \cite{WinskiSommerer2004a,Sakurai2007a}.

In the original model interaction 
between only two representative neurons is simulated: the orexin 
neuron and the local glutamate neuron that are reciprocally 
connected to each other according to the experimentally 
established physiological connections \cite{Li2002a}. Both orexin and 
glutamate neurons are firing during wakefulness and are silent 
during sleep. The transitions between firing and silence are 
governed by the interplay between the circadian input and 
homeostatic mechanisms as initially proposed by Borbely  \cite{Borbely1982}. 
For simplicity, in this model only a single type of orexin neurotransmitter 
(instead of the two types actually known) is considered, 
and it is assumed that the system can be either in the wake 
state or in a generic non-Rapid Eye Movement sleep 
state, without specifying ultradian structure of sleep.  
Also this model did not consider noise effects, and 
diversity could not be included since there are only 
two neurons present.

In the present paper we extend the above described 
two-neuron model to a more realistic multi-unit model 
with heterogeneous neurons. The aim of the study is 
to first of all investigate how the presence of diversity 
in the neuronal population affects sleep-wake transitions, 
since it is well-known that neurons are highly heterogeneous 
by their nature. In particular, within the orexin neurons 
population significant intrinsic diversity can be found: 
different electrophysiological properties, sizes in the 
diameter range $15$-$40~\mu\mathrm{m}$, and various shapes such as a spherical, 
fusiform, or multipolar  \cite{Peyron1998a,Eggermann2003a,Sakurai2007a}. 
Secondly, also stochastic fluctuations, representing current noise, are added to the model 
and the response of the system is studied for different levels 
of noise. The question naturally arises, to what extent noise 
and diversity are essential ingredients for the functioning of 
assemblies of neurons and other complex systems, 
and what is the optimal level of noise and diversity required for 
the emergence of an optimal response to external stimuli. It is 
shown below that the model under study presents both diversity-induced 
resonance and stochastic resonance, but the former appears more clear 
and robust, since it is always associated with a regular almost-periodic 
spiking-silence activity, rather than to the irregular random transitions 
characterizing the stochastic resonance regime.

\section{Materials and Methods}

In this section the two-neuron model of sleep-wake
cycles \cite{Postnova2009a} is described and some
examples of dynamics in the presence of an external periodic signal
are illustrated. 
Further, this model is extended to account for multiple neurons
dynamics and heterogeneity, and a simple quantitative criterion to
estimate the quality of a sleep-wake cycle is introduced.
This criterion will be used in the Results section to compare
sleep-wake cycles dynamics obtained at different parameter sets.

\begin{figure}[!ht]
\begin{center}
 \includegraphics[width=3in]{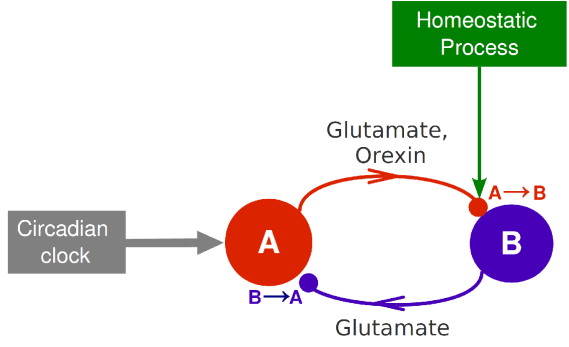}
\end{center} 
\caption{
  {\bf Scheme of the two-neuron model of the sleep-wake
    cycle \cite{Postnova2009a}.} 
  The $\mathrm{A}\to\mathrm{B}$ red arrow from the orexin-producing neuron \textrm{A} (red circle) to the
  neuron B (blue circle) represents the glutamate projection as well
  as the orexin projection regulating the homeostatic process.
  The blue arrow represents the $\mathrm{B}\to\mathrm{A}$ glutamate projection.
  The neuron \textrm{A} is also acted upon by a periodic signal representing
  the effect of the circadian clock.
  \label{scheme}
 }
\end{figure}

\subsection{The two-neuron model}
\label{2n-neurons}

The original model of the homeostatic regulation of sleep has a minimal
structure consisting of two representative interacting neurons A and
B, as depicted in Fig.~\ref{scheme}. 
The neuron A simulates a representative neuron from the orexinergic neuronal 
population, while the neuron B represents a local glutamate
interneuron (for details see  \cite{Postnova2009a}). 
The state of wakefulness or sleep is determined by the
firing regime of neurons A and B, since these
neurons are known to fire during wakefulness and be almost silent
during sleep (see e.g.  \cite{Sakurai2007a}). 

Interaction between the neurons A and B takes place through glutamate
and orexin neurotransmitters, as detailed below. 
The neuron A is acted upon by a stimulus in pace with the circadian rhythm,
here treated as a periodic external signal --- a simplification
justified by its independence from the homeostatic
process \cite{Borbely2000a}. The homeostatic process itself is described
by an additional macroscopic variable $M(t)$ simulating
availability of orexin.

Dynamics of the neurons A and B are based on a
Hodgkin-Huxley-type model  \cite{Hodgkin1952a}. 
The membrane potentials of the neurons A ($V_\mathrm{A}(t)$) and B
($V_\mathrm{B}(t)$) are thus calculated as:
\begin{eqnarray}
   C_\mathrm{A} \frac{dV_\mathrm{A}\!}{dt}
   &=&
    I_\mathrm{ext}
  + \xi_\mathrm{A}
  - I_\mathrm{A,L}
  - I_{\mathrm{A,Na}} 
  - I_{\mathrm{A,K}}
  - I_{\mathrm{A,gl}} 
  \label{VV0}
   \\
  &\equiv&
    I_\mathrm{ext}
  + \xi_\mathrm{A}
  - g_\mathrm{L} [V_\mathrm{A}\! \!-\! E_\mathrm{L}]
  - g_\mathrm{Na} [V_\mathrm{A}\! \!-\! E_\mathrm{Na}] a_\mathrm{A,Na}
   \nonumber \\
  &-& g_\mathrm{K} [V_\mathrm{A}\! \!-\! E_\mathrm{K}] a_\mathrm{A,K}
  - g_\mathrm{gl} [V_\mathrm{A}\! \!-\! E_\mathrm{gl}] a_\mathrm{A,gl}, \nonumber
  \\
   C_\mathrm{B} \frac{dV_\mathrm{B}}{dt}
  &=&
    \xi_\mathrm{B}
  - I_{\mathrm{B,L}}
  - I_{\mathrm{B,Na}}
  - I_{\mathrm{B,K}}
  - I_{\mathrm{B,gl}}
  - I_{\mathrm{ox}}
 \label{VV2}
\\
  &\equiv&
    \xi_\mathrm{B}
  - g_\mathrm{L} [V_\mathrm{B}\! \!-\! E_\mathrm{L}]
  - g_\mathrm{Na} [V_\mathrm{B}\! \!-\! E_\mathrm{Na}] a_\mathrm{B,Na}
   \nonumber \\
   - g_\mathrm{K} [V_\mathrm{B}\! \!&-&\! E_\mathrm{K}] a_\mathrm{B,K}
  - g_\mathrm{gl} [V_\mathrm{B}\! \!-\! E_\mathrm{gl}] a_\mathrm{B,gl}
  - g_\mathrm{ox} [V_\mathrm{B}\! \!-\! E_\mathrm{ox}] a_\mathrm{ox}, \nonumber
\end{eqnarray}
where $C_p$ ($p = \mathrm{A},\mathrm{B}$) are the membrane capacitances
per unit area of the respective neurons, $I_\alpha$ ($\alpha =
\mathrm{L},\mathrm{Na},\mathrm{K}, \mathrm{gl}, \mathrm{ox}$) are the
ionic currents, $g_\alpha$ are the maximum conductances, and
$E_\alpha$ are the equilibrium potentials.
The capacitance values are taken as 
$C_\mathrm{A} = C_\mathrm{B} = 1 \mu\mathrm{F/cm}^2$.
The values of all the other model parameters are listed in Table
\ref{table1}.

\begin{table*}[!ht]
\caption{\label{table1}
         {\bf Parameters of the two-neuron
           model \cite{Postnova2009a}.}}
\begin{tabular}{|l|l|l|l|l|l|}
\hline
      & Conductance
      & Equilibrium
      & Slope
      & Threshold
      & Time \\
      & 
      & Potential
      & Parameter
      & Potential
      & Scales\\
      & ($\mu\mathrm{S/cm}^2$)
      & ($\mathrm{mV}$)
      & ($\mathrm{mV}^{-1}$)
      & ($\mathrm{mV}$)
      & ($\mathrm{ms}$)\\
\hline
L (Leakage current)
      & $g_\mathrm{L} = 0.1$
      & $E_\mathrm{L} = -60$
      & 
      &
      & \\
\hline
Na (Sodium current)
      & $g_\mathrm{Na} = 3$
      & $E_\mathrm{Na} = 50$
      & $S_\mathrm{Na} = 0.25$
      & $W_\mathrm{Na} = -25$
      & ($\tau_\mathrm{Na} \approx 0)$\\
\hline
K (Potassium current)
      &$g_\mathrm{K} = 4$
      &$E_\mathrm{K} = -90$
      &$S_\mathrm{K} = 0.25$
      &$W_\mathrm{K} = -25$
      &$\tau_\mathrm{K} = 2$\\
\hline
gl (Glutamate current)
      &$g_\mathrm{gl} = 0.15$
      &$E_\mathrm{gl} = 50$
      &$S_\mathrm{gl} = 1$
      &$W_\mathrm{gl} = -20$
      &$\tau_\mathrm{gl} = 30$\\
\hline
ox (Orexin current)
      &$g_\mathrm{ox} = 0.135$
      &$E_\mathrm{ox} = 50$
      &$S_\mathrm{ox} = 1$
      &$W_\mathrm{ox} = -20$
      &$\tau_\mathrm{ox} = 300$\\
      &$g_\mathrm{ox} = 0.2$
      &
      &
      &
      &$\tau^{+}_\mathrm{ox} = 7500$\\
      &~~
      &~~ 
      &~~ 
      &~~ 
      &$\tau^{-}_\mathrm{ox} = 920$\\
\hline
Periodic current
      & 
      &
      & 
      & 
      &$\tau = 24000$\\
      & 
      &
      & 
      & 
      &$\tau_0 = 500$\\
\hline
\end{tabular}
\end{table*}

In the following we give a detailed explanation of different parts of the model.
\begin{itemize}

\item

  \emph{External forces}.
  The current $I_\mathrm{ext}$ acting on the neuron A and the noise
  currents $\xi_p(t)$, $p = \mathrm{A},\mathrm{B}$, can be
  considered as external forces, in the sense that they do not depend on the system variables.

  The \emph{external current} $I_\mathrm{ext}(t)$ is assumed to simulate a
  stimulus associated with the circadian rhythm. For simplicity in the present study a periodic
  pulse input is used to introduce circadian activation of the system:
  $\tau$, $I_\mathrm{ext}(t) = I_\mathrm{ext}(t+\tau)$.
  Such current can be interpreted as an awakening effect of an alarm clock 
  or some other disturbance coming with a period of 24 hours. 
  In the following we employ a train of rectangular pulses with
  length $\tau_0$ ($\tau_0 < \tau$) and height $I_0$, 
  as depicted in Fig.~\ref{fig_evolution}-top,
  \begin{eqnarray}
    I_\mathrm{ext}(t)~=~&I_0 \, , 
    ~~~~ &n\tau \le~t~< n\tau + \tau_0 \, ,
    \nonumber \\
    =~&0 \, , 
    ~~~~ &n\tau + \tau_0 \le~t~< (n+1)\tau \, ,
    \label{Idef}
  \end{eqnarray}
  where $n$ is an integer.
  This simple form is chosen because it is convenient for
  carrying out a systematic study of the neuron response at different
  parameters sets. However, it should be kept in mind that it represents a
  drastic simplification, and more realistic shapes of circadian
  currents can also be used  \cite{Postnova2009a}.

  The \emph{noise} term $\xi_p(t)$ represents fluctuating currents
  that are known to be always present in neurons. For simplicity, we assume zero-average Gaussian white-noise processes:
  \begin{eqnarray}
    \label{Xidef}
    &&\langle \xi_p(t) \rangle \!=\! 0; 
    \nonumber \\
    &&\langle \xi_p(t) \, \xi_{p'}(s) \rangle 
    \!=\! 2 \, D_p \, \delta_{p,p'} \, \delta(t\!-\!s),
   ~~ p, p' \!=\! \mathrm{A,B} ,
  \end{eqnarray}
with $D_p$ being the noise intensity.

\begin{figure*}[!ht]
\begin{center}
\includegraphics[width=6.8in]{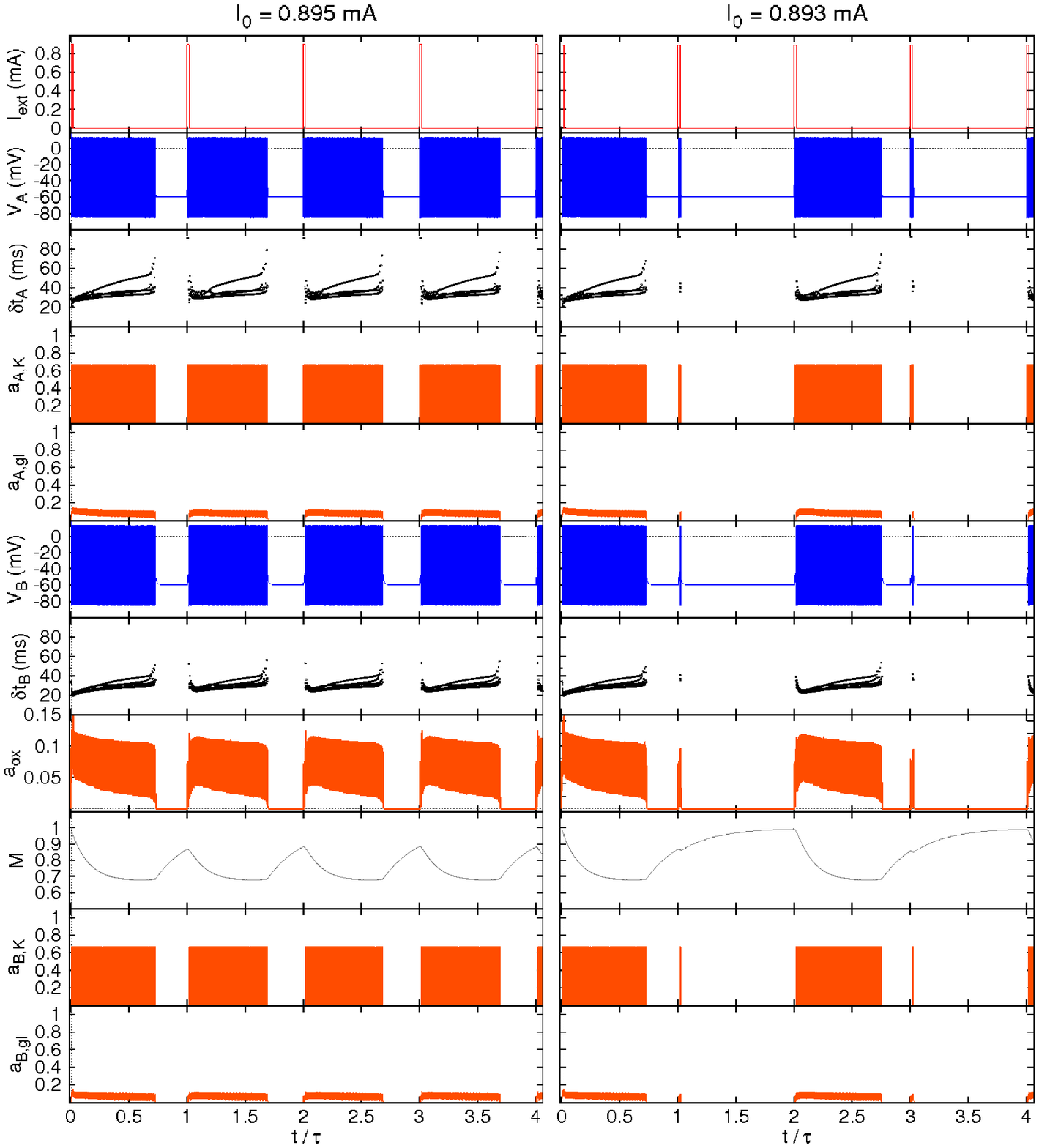}
\end{center}
 \caption{
  {\bf Response of the two-neuron model}.
  Main variables and inter-spike times $\delta t_\mathrm{p}$
  ($p = \mathrm{A}, \mathrm{B}$) \textit{versus} time for a pulse height
  $I_0 = 0.895~\mathrm{mA}$ (left) and $I_0 = 0.893~\mathrm{mA}$ (right),
  see text for details.
 \label{fig_evolution}
 }
\end{figure*}

\item

  \emph{Internal dynamics}.
  The leakage, sodium, and potassium currents 
  $I_{p,\alpha}$ 
  ($p = \mathrm{A}, \mathrm{B}$; 
   $\alpha = \mathrm{L}, \mathrm{Na}, \mathrm{K}$)
  in the equation of the neuron $p$ depend only on the variables of the
  \emph{same} neuron $p$ and, thus, describe the neuronal internal
  dynamics.

  The \emph{leakage currents} 
  $I_{p,\mathrm{L}} = g_\mathrm{L} (V_\mathrm{p} \!-\! E_\mathrm{L})$
  represent a flow of ions with a small conductance
  $g_\mathrm{L} \approx 0.1~\mu\mathrm{S/cm}^2$
  driving the membrane potential toward the negative value 
  $E_\mathrm{L} \approx -60~\mathrm{mV}$.

  The depolarizing \emph{Na-currents}
  $I_{p,\mathrm{Na}} = g_\mathrm{Na} (V_{p} \!-\! E_\mathrm{Na}) \, a_{p,\mathrm{Na}}$
  have a maximum conductance 
  $g_\mathrm{Na} = 3~\mu\mathrm{S/cm}^2$ 
  and a large positive equilibrium potential
  $E_\mathrm{Na} = 50~\mathrm{mV}$.
  The activation variables 
  $a_{p,\mathrm{Na}}(t)$, with $0 \le a_{p,\mathrm{Na}} \le 1$, 
  represent the fraction of open ion-channels contributing to the Na current.
  Because of their fast activation relative to the other time scales, 
  the Na-current is assumed to be activated instantaneously, according to its voltage-dependency:
  \begin{eqnarray}
  a_{p,\mathrm{Na}}  
  = 
  \Phi(S_\mathrm{Na} (V_p - W_\mathrm{Na})) \, ,
  \label{aNa}
  \end{eqnarray}
  where $\Phi(x)$ is the sigmoid function
  \begin{eqnarray}
  \Phi(x) = \frac{1}{ 1 + \exp(-x) } \, ,
  \label{Phi}
  \end{eqnarray}
  $S_\mathrm{Na}$ is the steepness of the sigmoid function
  and $W_\mathrm{Na}$ is the half-activation potential. 

  The repolarizing \emph{K-currents} 
  $I_{p,\mathrm{K}} = g_\mathrm{K} (V_{p} \! - \! E_\mathrm{K}) \, a_{p,\mathrm{K}}$
  are characterized by a maximum conductance
  $g_\mathrm{K} = 4~\mu\mathrm{S/cm}^2$,
  a large negative equilibrium potential $E_\mathrm{K} = -90~\mathrm{mV}$,
  and a longer activation time than the depolarizing Na-current, namely
  $\tau_\mathrm{K} = 2~\mathrm{ms}$.
  Consequently, the dynamics of the K-currents activation variables are modelled as
  \begin{eqnarray}
  \label{aK}
  \frac{da_{p,\mathrm{K}}}{dt} 
  &=& 
  - \, \frac{1}{\tau_\mathrm{K}} 
  \left[
  a_{p,\mathrm{K}} - \Phi(S_\mathrm{K} (V_p - W_\mathrm{K}))
  \right] \, ,
  \end{eqnarray}
  where $\Phi(x)$ is defined in Eq.~(\ref{Phi}).

  \emph{Couplings}.
The neurons A and B are mutually coupled by chemical synapses through the glutamate-induced 
($I_{\mathrm{p,gl}}$) and the orexin-induced ($I_\mathrm{ox}$) currents. 
Unlike the Na and K currents, $I_{\mathrm{p,gl}}$ and $I_\mathrm{ox}$ 
depend on the activity of both presynaptic and postsynaptic neurons. 
The activation variables $a_{\mathrm{p,gl}}$ and $a_\mathrm{ox}$ depend 
on the appearance of a spike in the presynaptic neuron, i.e. on the presynaptic 
voltage. Additionally these currents  depend on the voltage of the postsynaptic 
neuron, similarly to other ionic currents. Both glutamate and orexin are excitatory 
neurotransmitters, so they are assumed to open depolarizing ion channels, such as Na-channels. 

The activations of the glutamate-induced currents are modeled as: 
\begin{eqnarray}
\label{Igl}
  \label{agl}
  \frac {da_{\mathrm{p,gl}}} {dt} 
  &=& 
  - \frac{1}{\tau_\mathrm{gl}} 
     \left[\,
        a_{\mathrm{p,gl}} 
      - \Phi(S_\mathrm{gl} (V_{\bar{p}} \!-\! W_\mathrm{gl}))
  \,  \right] 
  \, ,
   \nonumber \\
  &&\bar{p} \!=\! \mathrm{B~~if~~} p = \mathrm{A};~~\bar{p} = \mathrm{A~if~~} p = \mathrm{B} .
\end{eqnarray}
This equation is similar to Eq.~(\ref{aK}) but has the important
difference that the equilibrium value 
$\Phi(S_\mathrm{gl} (V_{\bar{p}} \!-\! W_\mathrm{gl}))$
for the activation variable $a_{\mathrm{p,gl}}$ 
depends on the membrane  potential 
$V_{\bar{p}}$ of the \emph{other} neuron $\bar{p}$
($\bar{p} \!=\! \mathrm{B}$ if $p = \mathrm{A}$, 
$\bar{p} = \mathrm{A}$ if $p = \mathrm{B}$).
The time constant $\tau_\mathrm{gl} = 30~\mathrm{ms}$
accounts for the delay coming from the activation of glutamate
receptors, and the following activation of ion channels.

The \emph{orexin-induced current} represents the effect of orexin produced 
by the neuron A and acting on the neuron B. It is modeled in a form similar to the glutamate-induced current. 
This current provides a simplified description of the effects of orexin
on the neuron B which appear after a complex series of processes,
involving production of orexin in the soma of the neurons, its release in the synaptic cleft, 
and activation of G-protein coupled metabotropic receptors.
The dynamics of the activation variable $a_\mathrm{ox}(t)$
depend not only on the membrane potential $V_\mathrm{A}(t)$, but
are also related to the availability of orexin at time $t$, described
by the additional variable $M(t)$ ($0 \le M \le 1$).
The dynamics of the variables $a_\mathrm{ox}(t)$ and
$M(t)$ are defined by the equations:
\begin{eqnarray}
  \label{bor}
  \frac{da_\mathrm{ox}}{dt}
  \!=\! 
  &-& \frac{1}{\tau_\mathrm{ox}} 
  [
    a_\mathrm{ox} 
  \!-\! M \times
   \Phi(S_\mathrm{ox} (V_\mathrm{A} \!-\! W_\mathrm{ox}))
  ] \, ,
  \\
  \frac{dM}{dt}
  \!=\! 
  &-& \frac{1}{\tau^{+}_\mathrm{ox}} (M \!-\! 1) 
  \nonumber \\
    &-& \frac{1}{\tau^{-}_\mathrm{ox}} 
    M \times \Phi(S_\mathrm{ox} (V_\mathrm{A} \!-\!
    W_\mathrm{ox})) .~~~~
  \label{M}
  \end{eqnarray}
The term $M \times \Phi(S_\mathrm{ox} (V_\mathrm{A} \!-\! W_\mathrm{ox}))$ 
in the Eq.~(\ref{bor}) reflects activation of the synaptic current due to 
appearance of a spike in the presynaptic neuron A. At the same time it 
determines the rate of orexin availability reduction in Eq.~(\ref{M}) due 
to spiking of the neuron A with a time constant $\tau^{-}_\mathrm{ox}$. 
The first term in Eq.~(\ref{M}) determines recovery rate of the orexin 
availability with time constant $\tau^{+}_\mathrm{ox}$.

The meaning of the product 
$M(t) \times \Phi(S_\mathrm{ox} (V_\mathrm{A} \!-\! W_\mathrm{ox}))$
is that there is orexin-induced activity in neuron B if
(1) there is enough orexin available above a critical threshold 
[$M(t) > M_{critical}$], and 
(2) the neuron  A is in the firing state [$\Phi(t) \approx 1$].

The time constants $\tau^{\pm}_\mathrm{ox}$ accounting for the orexin dynamics  
are much longer than the time constants associated with ionic current terms.
The time constant $\tau_\mathrm{ox}$ of the homeostatic
regulation process is even longer, being of the order of
magnitude of the daily period $\tau$.

For numerical convenience, simulations are made over
rescaled daily and orexin time scales:
the daily period was assumed to be $\tau = 24~\mathrm{s}$,
instead of $\tau=24~\mathrm{h}$, achieved through a suitable rescaling, which was applied
to the orexin time scale $\tau_\mathrm{ox}$ and the production and
reduction times $\tau^{\pm}_\mathrm{ox}$.
The other time parameters are left unchanged.
Since such rescaled $\tau_\mathrm{ox}$ and $\tau^{\pm}_\mathrm{ox}$
are still much larger than any other time scale of the microscopic
dynamics, the rescaling does not change the main results
of the simulations. See  \cite{Postnova2009a} for a detailed
validation of such rescaling procedure.

\end{itemize}

All the parameter values for the currents are listed in Table \ref{table1}.
It is assumed that the neurons A and B share the same parameter values, unless  specified otherwise. 
Such an assumption is justified, because the major properties of these
neurons required for the model are the tonic firing (periodic single
spike activity) and silent states. Without any external input both
neurons should be in a silent state, while they are brought to firing
activity in response to depolarization. Therefore, change of
parameters in a physiologically allowed range would primarily lead to
the different amount of depolarization needed to excite neurons, and
would not affect the major outcomes of the simulations.


The system defined above is essentially an excitable feedback system,
i.e. both the external input of sufficient strength and the AB coupling are essential elements
for maintaining firing activity of the neurons. Orexin-related dynamics, with the associated long time scales,
are expected to direct the homeostatic sleep process, which regulates the sleep-wake transitions. 
The healthy sleep-wake cycles in this system are realized as follows:
\begin{itemize}
\item
\emph{Initiating wakefulness}. 
A sufficiently strong or long external signal or a stimulus associated to the
circadian rhythm,  e.g. the idealized rectangular pulse
considered here, activates the system and
induces firing activity in the neuron A. Due to the excitatory
synaptic connection from A to B, the neuron B is also activated. 
\item
\emph{Maintaining wakefulness}.
Once the pulse is finished and the external current is zero, the system
remains in the wake state (i.e. both neurons A and B are firing) 
due to reciprocal excitation between the neurons. The firing activity 
lasts for a fraction $q$ of the daily period $\tau$. Ideally one can assume $q = 2/3$, 
corresponding to 16 hours for a day of 24 hours, i.e. 16 seconds for the daily period 
$\tau = 24~\mathrm{s}$ with the time scales of the model considered here. 
\item
\emph{Initiating sleep}.
The firing stops in both neurons due to decreased availability of orexin according 
to the dynamics of $M(t)$. This is associated witch the transition from wake to sleep. 
\end{itemize}

Two examples of the two-neuron model dynamics without noise are
illustrated in Fig.~\ref{fig_evolution}. The left part of the figure
represents the response obtained for a
pulse length $\tau_0 = 500~\mathrm{ms}$ 
and height $I_0 = 0.895~\mu\mathrm{A/cm}^2$.
In each period orexin is depleted during the neuronal activity and 
recovered  while the neurons are silent. The stimulus parameters used
in this example have been intentionally chosen close to the critical
firing threshold, so that by slightly reducing the pulse height or
length, the periodic appearance of a continuous time interval of
spiking regime is lost. Such case is demonstrated in the right hand
side of the figure, where the current pulse height is slightly lower,
$I_0 = 0.893~\mathrm{mA}$, while all the other parameters are kept the
same. There the prolonged wake state is induced only every other day,
because the input is insufficient to induce sustained spiking at the
same levels of orexin availability. By reducing the pulse amplitude or
duration even further it is possible to observe different behaviors
such as triple or higher-order periodicities.

\begin{figure}[!ht]
\begin{center}
\includegraphics[width=2.5in]{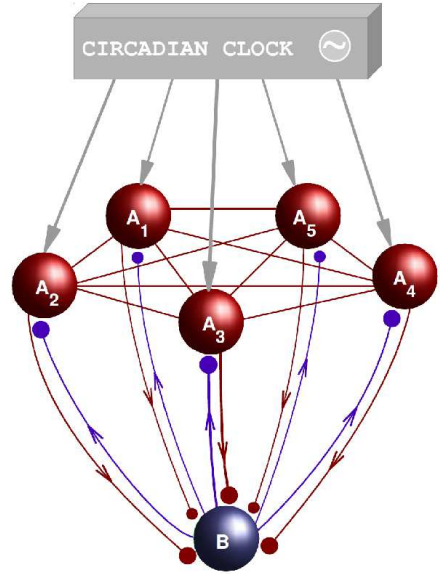}
\end{center}
 \caption{
 {\bf Scheme of the heterogeneous model}.
 Example of model system with $N_\mathrm{A} = 5$ orexin-producing
 neurons $\{\mathrm{A}_i\}$, $i = 1, \dots N_\mathrm{A}$ (red spheres)
 and one neuron B (blue sphere). 
 The neurons A interact with each other through an all-to-all coupling
 (red lines).
 Blue and red projections have a meaning similar to those of
 Fig.~\ref{scheme}:
 the neuron B is coupled to the neurons $\mathrm{A}_i$ through
 parallel glutamate projections,
 while each neuron $\mathrm{A}_i$ is coupled to neuron B through a
 glutamate and an orexin projection. 
 The neurons A are also acted upon by a stimulus representing the
 effect of the circadian clock (gray arrows). 
 \label{fig_N-neurons-2}
 }
\end{figure}

\subsection{The heterogeneous model}
\label{N-neurons}
As a step toward a more realistic model we generalize the two-neuron model into a 
heterogeneous multi-neuron model. For simplicity we first increase the number of orexin neurons only. 
To do this we replace the single neuron A by a 
set of $N_\mathrm{A}$ neurons $\{\mathrm{A}_i\}$  ($i = 1, \dots, N_\mathrm{A}$), while
still maintaining only one neuron B. Also, in this paper we assume that 
the diversity is constant in time in order to consider the simplest case possible. 

In reality a certain level of heterogeneity is observed in all neuronal parameters. However, given that
our model neurons are simple pacemaking neurons such diversification of different 
model parameters (in a physiologically allowed range) would simply lead to slightly 
different firing rates of the neurons. This, in turn, will result in diversity in 
activations of synaptic currents, which can be mimicked by simply diversifying their 
activation thresholds. Thus, in the following we can limit ourselves to studying the 
effects of diversity in activation thresholds of synaptic currents without loss of generality. 
Furthermore, as a first step, the heterogeneity is only introduced in the glutamate-induced 
currents to avoid having a too complicated system, which would become difficult to understand.

With regard to the coupling topology among the orexin neurons, 
so far there is no detailed experimental data. Therefore, for simplicity, 
we chose an all-to-all coupling via gap junctions, but other variations 
can be tested in the future. The intensity of the coupling has been chosen large enough to
ensure that the neurons $\mathrm{A}_i$ respond in pace to the external current. The equations 
of the two-neuron model are modified accordingly.

\begin{itemize}

\item
\emph{Dynamics of the neurons $\mathrm{A}_i$}. 
The membrane potentials $V^{\,(i)}_\mathrm{A}\!(t)$ of the neurons $\mathrm{A}_i$, 
are described by equations analogous to Eq.~(\ref{VV0}):
\begin{eqnarray}
 &&C_\mathrm{A} \frac{dV^{\,(i)}_\mathrm{A}\!}{dt}
 \nonumber\\
 &&= 
    I_\mathrm{ext}
  + \xi_\mathrm{A}^{(i)}
  - I^{\,(i)}_\mathrm{A,L} 
  - I^{\,(i)}_\mathrm{A,Na} 
  - I^{\,(i)}_\mathrm{A,K}
  - I^{\,(i)}_\mathrm{A,gl} 
  - \sum_j I_{ij} 
 \nonumber\\
  &&=
       I_\mathrm{ext}  
     + \xi_\mathrm{A}^{(i)}
    -  g_\mathrm{L}  
       [V^{\,(i)}_\mathrm{A}\! \!-\! E_\mathrm{L}]
    -  g_\mathrm{Na} 
       [V^{\,(i)}_\mathrm{A}\! \!-\! E_\mathrm{Na}] 
       \, a^{\,(i)}_\mathrm{A,Na}
  \nonumber \\
  &&-  g_\mathrm{K} 
       [V^{\,(i)}_\mathrm{A}\! \!-\! E_\mathrm{K}] 
       \, a^{\,(i)}_\mathrm{A,K}
    - g_\mathrm{gl} 
      [V^{\,(i)}_\mathrm{A}\! \!-\! E_\mathrm{gl}] 
      \, a^{\,(i)}_\mathrm{A,gl}
  \nonumber \\
  &&- k_\mathrm{\,int} 
      \sum_j [ V^{\,(i)}_\mathrm{A} - V^{\,(j)}_\mathrm{A} ] 
  \, .
  \label{VV3}
 \end{eqnarray}
The current terms are similar to those in the two-neuron model, 
apart from the additional coupling currents between two generic
neurons $\mathrm{A}_i$ and $\mathrm{A}_j$, 
$I_{ij} = k_\mathrm{\,int}(V^{(i)}_\mathrm{A} \!-\! V^{(j)}_\mathrm{A})$, 
with
$i,j = 1, \dots, N_\mathrm{A}$,
where $k_\mathrm{\,int}$ is the gap junctions conductance that can be
treated as coupling strength. The currents' activation variables
$a^{\,(i)}_\mathrm{A,Na}$ and $a^{\,(i)}_\mathrm{A,K}$ are modeled in
accord with the equations of the two-neuron model. Note that the
specific values of the activation variables will be different for
different neurons since they depend on voltages of each particular
neuron $\mathrm{A}_i$.

For simplicity the same external current $I_\mathrm{ext}(t)$  given
by Eq.~(\ref{Idef}) is assumed to act on all neurons $\mathrm{A}_i$ 
(see Fig.~\ref{fig_N-neurons-2}).
The noise terms $\xi_\mathrm{A}^{(i)}(t)$ as well as the noise 
$\xi_\mathrm{B}(t)$ acting on the neuron B (see below) are also
defined similarly and assumed to be statistically independent from each
other. For convenience the properties of all stochastic forces are 
written together ($i,j = 1, \dots, N_\mathrm{A}$):
\begin{eqnarray}
  \label{Xidefi}
  &&
  \langle \xi_\mathrm{A}^{(i)}(t) \rangle 
  =
  \langle \xi_\mathrm{B}(t) \rangle 
  = 
  0 
  \, ,
  \nonumber \\
  &&
  \langle 
    \xi_\mathrm{A}^{(i)}(t) \, 
    \xi_\mathrm{B}(s) 
  \rangle  = 0 ,
  \nonumber \\
  &&\langle 
    \xi_\mathrm{A}^{(i)}(t) \, 
    \xi_\mathrm{A}^{(j)}(s) 
  \rangle 
  \!=\! 
  2 \, D_\mathrm{A} \, \delta_{i,j} \, \delta(t\!-\!s)
  \, ,
  \nonumber \\
  &&\langle \xi_\mathrm{B}(t) \, \xi_\mathrm{B}(s) \rangle 
  \!=\! 2 \, D_\mathrm{B} \, \delta(t\!-\!s) 
  \, .
  \end{eqnarray}

\item
\emph{Connections from the neuron \textrm{B} to the neurons $\mathrm{A}_i$}. 
The neuron B has glutamatergic synaptic inputs to each of the neurons $\mathrm{A}_i$ 
as depicted in Fig.~\ref{fig_N-neurons-2}. 
Diversity is introduced in the activation thresholds of the 
glutamate-induced currents according to the following equation for the activation variables:
\begin{eqnarray}
  \label{agl-N}
  \frac{da^{\,(i)}_\mathrm{A,gl}}{dt} 
  = - \frac{1}{\tau_\mathrm{gl}} 
  \left[
    a^{\,(i)}_\mathrm{A,gl} 
  - \Phi(S_\mathrm{gl} (V_\mathrm{B} \!-\! W^{(i)}_\mathrm{B,gl}))
  \right] .
  \end{eqnarray}
The thresholds $W^{\,(i)}_\mathrm{B,gl}$ adopt different values for each neuron
$\mathrm{A}_i$ that are independently extracted from a probability distribution defined later in the text.

\item
\emph{Connections from the neurons $\mathrm{A}_i$ to the neuron \textrm{B}}.
Each of the neurons $\mathrm{A}_i$ has synaptic projections to the neuron B. 
This is translated in the model by replacing the single glutamate- 
and orexin-induced currents with their averages such that Eqs.~(\ref{agl}) and
(\ref{bor}) for the activation variables become:

\begin{eqnarray}
  \label{bgl-N}
  &&\frac{da_\mathrm{B,gl}}{dt} 
  \nonumber \\
  &&= 
  - \frac{1}{\tau_\mathrm{gl}} 
  \! \left[
    a_\mathrm{B,gl} 
  - \frac{1}{N_\mathrm{A}} \sum_{i=1}^{N_\mathrm{A}} 
    \Phi(S_\mathrm{gl} (V^{\,(i)}_\mathrm{A} \!-\! W^{\,(i)}_\mathrm{A,gl}))
  \right] \! \! ,
  \\
  \label{bor-N}
  &&\frac{da_\mathrm{B,ox}}{dt} 
  \nonumber \\
  &&= 
  \! - \frac{1}{\tau_\mathrm{ox}} 
  \!\! \left[
    a_\mathrm{B,ox} 
  \! - \frac{1}{N_\mathrm{A}} \!\! \sum_{i=1}^{N_\mathrm{A}} \! M^{(i)} 
  \Phi(S_\mathrm{ox} (V^{\,(i)}_\mathrm{A} \!\!\!-\! W_\mathrm{ox}))
  \!\right]
  \end{eqnarray}

Note that diversity is again introduced in the activation 
thresholds of the glutamate-induced currents $W^{\,(i)}_\mathrm{A,gl}$ 
corresponding to heterogeneous $(\mathrm{A}_i\!\!\to\!\!\mathrm{B})$ synapses located at the neuron B. 
Due to the differences in the $\mathrm{A}_i$ neurons, the orexin availability function $M^{(i)}(t)$ is 
different for different neurons, although still following Eq.~(\ref{M}).

\end{itemize}


The above described set of equations constitutes the multi-neuron 
heterogeneous model of the homeostatic regulation of sleep. 
Numerical results were obtained using a variation of the Runge-Kutta
2nd-order method, which is suitable for equations with stochastic
terms, namely the Heun method \cite{SanMiguel2000a}.
Identical initial conditions were assumed for all neurons,
corresponding to a silent state.

\subsection{Quantifying the quality of the sleep-wake cycle}
\label{encoding}
In this section a heuristic criterion is introduced in order to evaluate and compare the quality
of the system responses obtained for different external signals or
internal parameter values.

For this purpose, the period $\tau$ is divided
into a ``day'' wakefulness sub-period of length $\tau_1 = q\tau$ 
and a ``night'' sleep sub-period of length 
$\tau_2 = (1-q) \tau$, with $\tau = \tau_1 + \tau_2$.
The quantity $q$ is defined as a \textit{wake fraction}.
A typical sleep-wake cycle with an eight-hour sleep sub-period has 
$q = 2/3$.
For the day corresponding to the $n$-th period 
$(n\tau, \!(n \!+\! 1)\tau)$,  
the ``day'' is represented by the sub-interval
$(n\tau, n\tau + \tau_1) = (n\tau, (n + q)\tau)$,
which covers the first fraction $q$ of the period,
while the ``night'' extends in the complementary fraction $(1-q)$ of
the period in the time interval 
$(n\tau \!+\! \tau_2, \!(n \!+\! 1)\tau) 
\!=\! 
((n \!+\! q)\tau, \!(n \!+\! 1)\tau)$.

For each period $n = 0, 1, \dots$, we compute wakefulness time
intervals $\Delta t_n^{(1)}$ and $\Delta t_n^{(2)}$
spent by the system in the wake state during the day,
$\Delta t_n^{(1)}$, and night, $\Delta t_n^{(2)}$.
The wake/sleep state is identified with the spiking/silent
regime. A simple quantitative estimate of the quality of the sleep-wake cycle
can, thus, be done through the following linear function of the wakefulness
time intervals,
\begin{eqnarray}
  \label{r2}
  r(\Delta t^{(1)}, \Delta t^{(2)}) 
  &=& 
    \frac{\Delta t^{(1)}}{\tau_1} 
    -
    \frac{\Delta t^{(2)}}{\tau_2}
  \, ,
\end{eqnarray}
where 
$\Delta t^{(\alpha)} = \sum_n \Delta t_n^{(\alpha)}/N_\mathrm{sp}$,
$\alpha=1,2$, 
represent the average of the wakefulness time intervals during the day
($\alpha=1$) and during the night ($\alpha=2$), with
$N_\mathrm{sp}$ being the total number of periods of the simulation.

The fractions $\Delta t^{(\alpha)}/\tau_\alpha$
($\alpha = 1,2$) can vary in the interval $(0,1)$;
then the coefficient $r$ in Eq.~(\ref{r2}) 
is limited in the interval $(-1,1)$.
The maximum value $r = 1$ corresponds to an optimal cycle with
$\Delta t^{(1)} = \tau_1$ (wakefulness during the entire day)
and $\Delta t^{(2)} = 0$ (sleep during the entire night);
any deviation from the optimal state ($r= 1$) comes either
from values $\Delta t^{(1)} < \tau_1$ (implying some sleep during the day) 
or values $\Delta t^{(2)} > 0$ (meaning at least some wakefulness in the night).
See the supporting information in the Appendix for further details on the definition of the
time intervals $\Delta t^{(1)}$, $\Delta t^{(2)}$ and the coefficient $r$.

\begin{figure}[!ht]
\begin{flushleft}
\includegraphics[width=9cm]{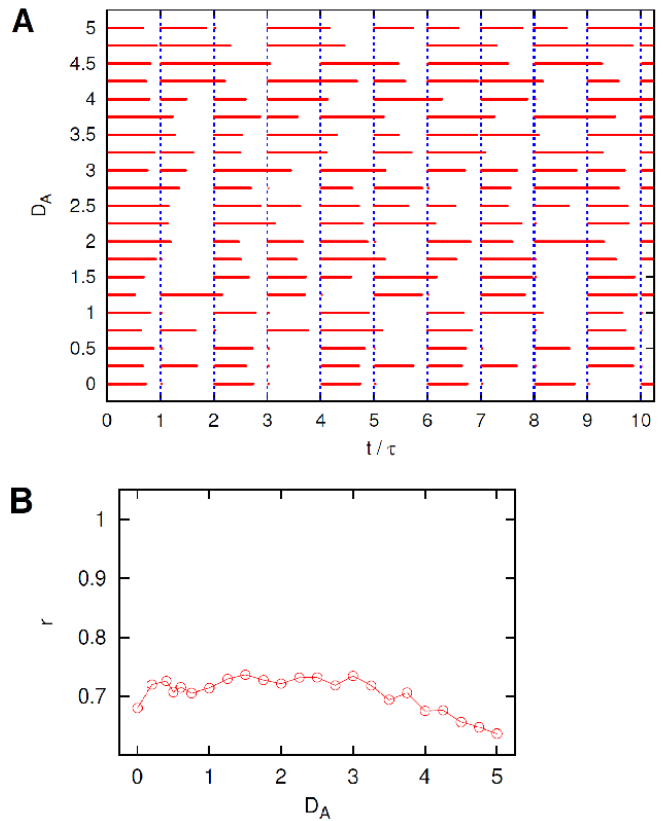}
\end{flushleft}
 \caption{
 {\bf Effect of noise in neurons $\{\mathrm{A}_i\}$}. 
 (A). Ten periods of the raster plots of neuron B for
 different intensities $D_\mathrm{A}$ of the noise acting on neurons
 $\{\mathrm{A}_i\}$.
 Vertical dashed lines mark the beginning of the pulses of the
 external current $I_\mathrm{ext}$, see text for details.
 (B). Coefficient $r$, from Eq.~(\ref{r2}),
 {\it versus} current noise intensity $D_\mathrm{A}$.
 \label{fig_rasterB-vs-DA}
 }
\end{figure}

\begin{figure*}[!ht]
\begin{center}
\includegraphics[width=7in]{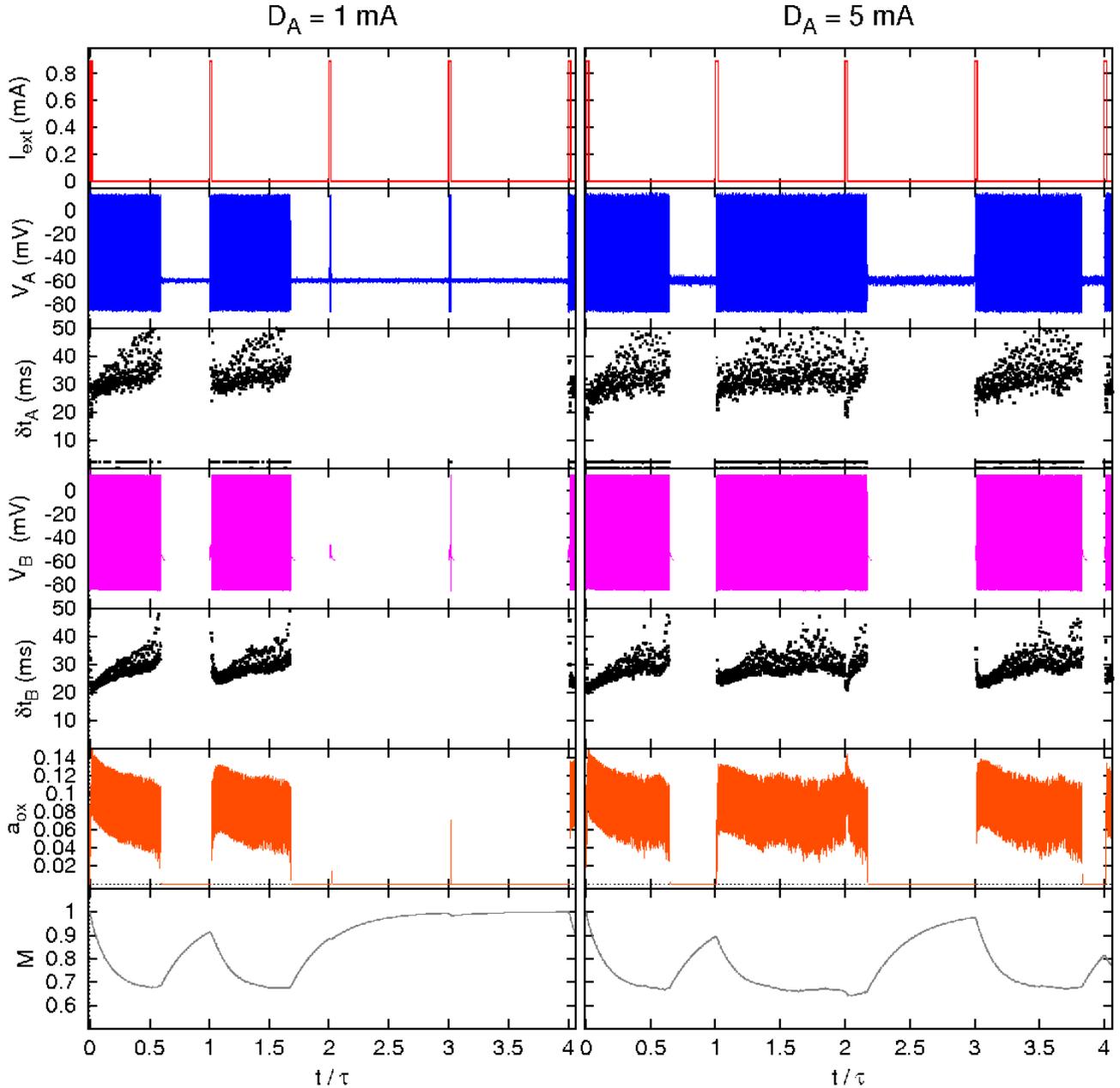}
\end{center}
 \caption{
  {\bf Effect of noise in neurons $\{\mathrm{A}_i\}$}. 
  Sample of four periods of some relevant variables and inter-spike
  times $\delta t_\mathrm{p}$ ($p = \mathrm{A}, \mathrm{B}$)
  \textit{versus} time of neuron $\mathrm{A}_1$ and neuron B for an
  intensity of noise in neurons A $D_\mathrm{A} = 1~\mathrm{mA}$ (left) and
  $D_\mathrm{A} = 5~\mathrm{mA}$ (right).
  Compare Fig.~\ref{fig_rasterB-vs-DA} and see text for details.
 \label{fig_evolution-DA}
 }
\end{figure*}

\section{Results}
\label{disorder}

In this section we study how the presence of disorder affects the
system response and discuss the main differences compared to the two-neuron model.
The term ``disorder'' is used here to refer to either
\textit{noise}, i.e. disorder in time (stochastic terms in the external
current), or \textit{diversity}, i.e a
quenched heterogeneity in the neuronal parameters.
These two aspects are studied separately. For the sake of simplicity,
we examine the response of the system to a periodic stimulus
represented by a train of short rectangular pulses as defined in
Materials and Methods.

In each of the examples considered, the initial configuration in
the absence of noise and diversity is the same as the sub-threshold state
illustrated in Fig.~\ref{fig_evolution}-right with a double-periodic response.
It is obtained for a reduced height of the current pulse 
$I_0 = 0.893~\mathrm{mA}$, while
the other parameters are unchanged as given in Table \ref{table1}.
The reason for starting from such an under-threshold non-optimal
configuration is that it is most sensitive and, thus, best illustrates
the effects of added noise or heterogeneity.
While a response with a double-periodicity may seem unrealistic, this
starting configuration is intended to be an example of non-optimal
response rather than a standard reference state.
In fact, in realistic situations noise and heterogeneity are always
present so that such a state without noise or diversity represents a
hypothetical system that would be obtained if one could switch off
noise or replace heterogeneous synapses with perfectly identical ones.
The results presented below suggest that a multi-periodic sleep-wake cycle
can be turned into a regular (single-periodic) one by adding a
suitable degree of disorder.

\subsection{Effect of noise}
\label{noise}

Here we investigate the effects of the noise currents
in the equations for the membrane potentials.
For clarity only the cases in which noise currents are present either 
in the neurons $\mathrm{A}_i$ or in the neuron B are considered.

\subsubsection{Noise in the neurons $\mathrm{A}_i$}

To study the effects of the noise currents $\xi_\mathrm{A}^{(i)}(t)$,
$i=1,\dots,N_\mathrm{A}$ acting only on the neurons $\mathrm{A}_i$
(as per Eq.~(\ref{VV3})) we set $D_\mathrm{B} = 0$.
Also, no diversity in the characteristic parameters of neurons
$\mathrm{A}_i$ is introduced.
We have simulated a system with $N_\mathrm{A} = 20$ identical
neurons and a single B neuron on a time interval $t \in (0, 100 \,\tau)$.
A raster plot for the activity of the neuron B at different values of $D_\mathrm{A}$
(indicated on the left) is shown in Fig.~\ref{fig_rasterB-vs-DA}-A.
The plot shows that
\begin{itemize}
\item 
  for small $D_\mathrm{A} \approx 0$ the system's configuration
  corresponds to the assumed non-optimal double-periodic solution;
\item 
  the system's response becomes slightly more regular and periodic as
    $D_\mathrm{A}$ is increased, 
  despite the fact that the neuron cannot initiate
  a firing event at the beginning of each period;
\item 
  as $D_\mathrm{A}$ becomes even larger the 
  neuron B keeps firing tonically for a longer and longer time interval
  (even longer than a single period) thus deteriorating the
  general quality of the response.
  
\end{itemize}
A sample of time dependence of the main variables in the interval
$(0, 4 \, \tau)$ for 
$D_\mathrm{A} = 1~\mathrm{mV}$ 
and $D_\mathrm{A} = 2~\mathrm{mV}$ is
illustrated in Fig.~\ref{fig_evolution-DA}.
In general, the type of variability induced by noise currents acting on the
neurons $\mathrm{A}_i$ affects both the firing initiation and, especially, its duration.
However, it  is difficult to establish an actual improvement of the
quality of such a response as a function of the noise intensity
$D_\mathrm{A}$, as even the coefficient $r$, shown in
Fig.~\ref{fig_rasterB-vs-DA}-B, suggests only a mild stochastic
resonant behavior characterized by a wide plateau at intermediate values
of $D_\mathrm{A}$.

\begin{figure}[!ht]
\begin{flushleft}
\includegraphics[width=9cm]{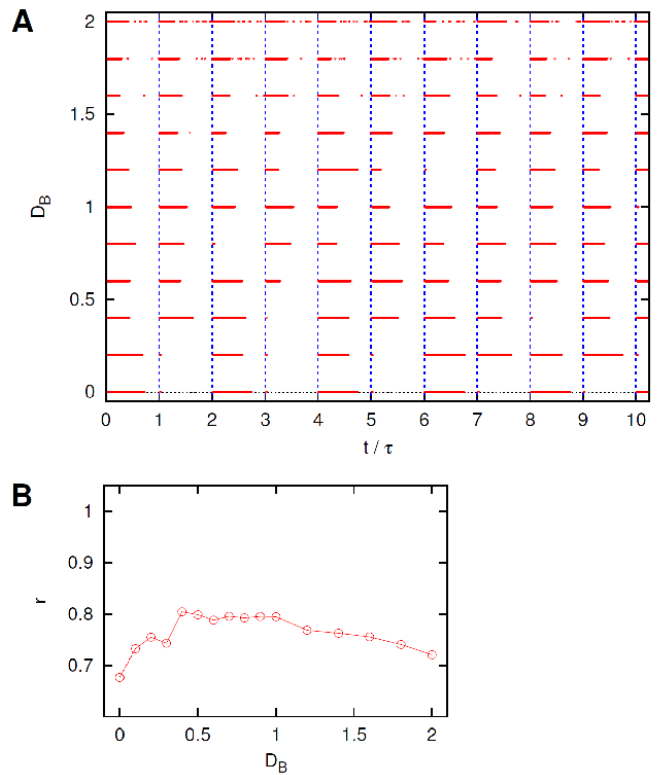}
\end{flushleft}
 \caption{
 {\bf Effect of noise in neuron B}.
 (A). Sample of ten periods of the raster plots of neuron B for different
 values of the intensity $D_\mathrm{B}$ of the noise acting on neuron B.
 Vertical dashed lines mark the beginning of the pulses of the
 external current $I_\mathrm{ext}$, see text for details.
 (B).  Quality of the sleep-wake cycle from the coefficient $r$, Eq.~(\ref{r2}), 
 \textit{versus} current noise intensity $D_\mathrm{B}$.
 \label{fig_rasterB-vs-DB}
 }
\end{figure}

\subsubsection{\bf Noise in the neuron \textrm{B}}

Here we consider the complementary case, in which $D_\mathrm{A} = 0$
and a current noise only affects the neuron B.
A sample of raster plots of the membrane potential of the neuron B 
is depicted in Fig.~\ref{fig_rasterB-vs-DB}-A in the time window $t \in (0, 10 \, \tau)$
for the values of the noise intensity $D_\mathrm{B}$
indicated on the vertical axis.

\begin{figure*}[!ht]
\begin{center}
\includegraphics[width=7in]{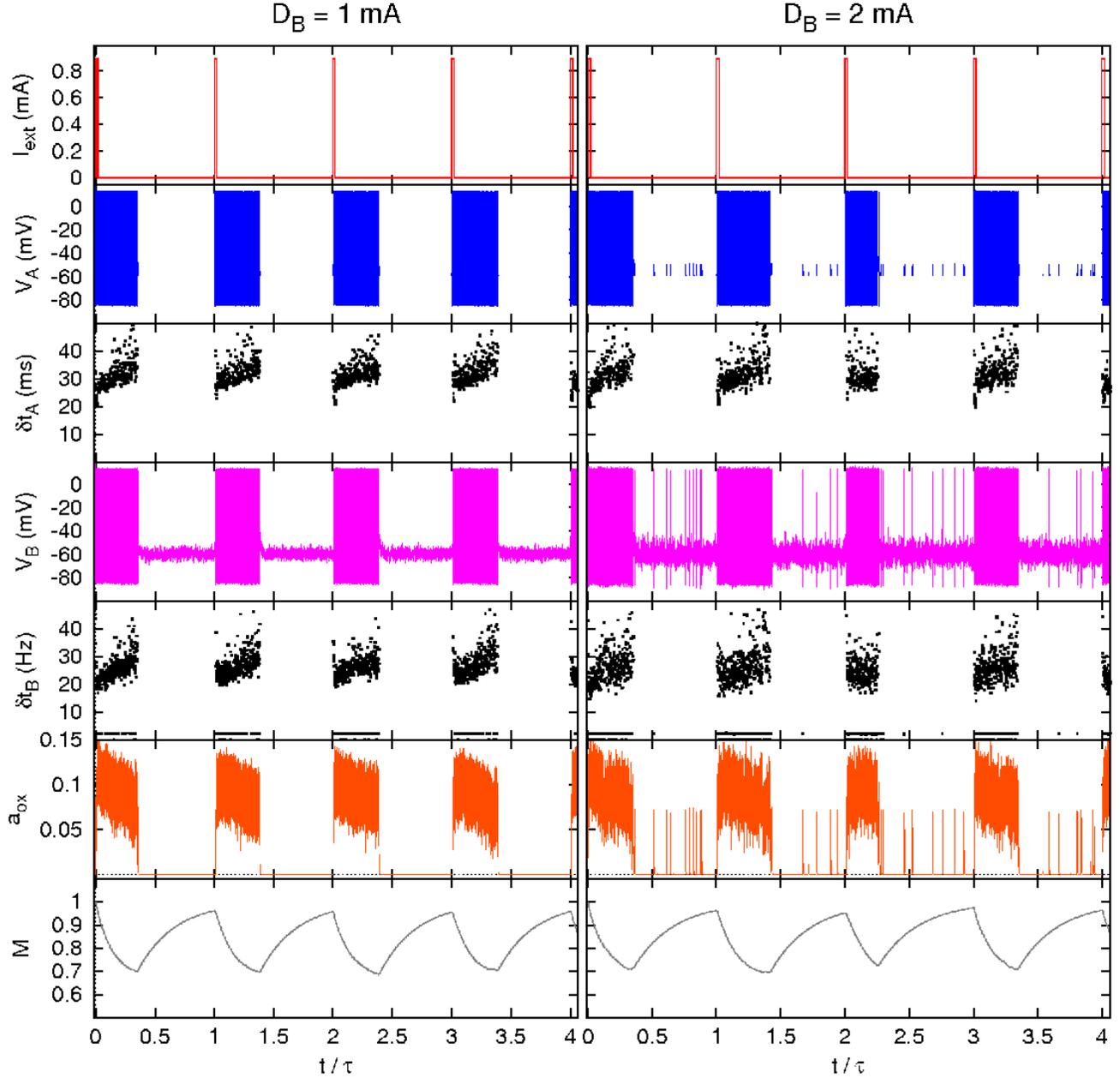}
\end{center}
 \caption{
  {\bf Effect of noise in neuron \textrm{B}}.
  Sample of four periods for some relevant variables and 
  inter-spike times $\delta t_\mathrm{p}$ ($p = \mathrm{A}, \mathrm{B}$)
  \textit{versus} time for neuron $\mathrm{A}_1$ and neuron B for an intensity
  of the noise acting on neuron B $D_\mathrm{B} = 1~\mathrm{mA}$ (left) and
  $D_\mathrm{B} = 2~\mathrm{mA}$ (right).
  Compare Fig.~\ref{fig_rasterB-vs-DB} and see text for details.
 \label{fig_evolution-DB}
 }
\end{figure*}

The raster plot in Fig.~\ref{fig_rasterB-vs-DB}-A indicates that:
\begin{itemize} 
\item 
  the smallest values of noise intensity $D_\mathrm{B} \approx 0$ correspond to the
  double-periodic configuration discussed above;
\item 
  the response becomes periodic, and the length of
  the firing periods more regular for higher values of
  $D_\mathrm{B}$;
\item 
  at larger values of $D_\mathrm{B}$ the state of sleep is
  frequently interrupted by almost isolated spikes at random times.
\end{itemize}
A representative example of time dependence of selected variables of the neurons
$\mathrm{A}_1$ and B are shown in Fig.~\ref{fig_evolution-DB}.
Note the different type of behavior induced by a high levels of noise
acting on the neuron B, compared to the case in which noise acts on the neurons
$\mathrm{A}_i$. In the first case irregular switching between the firing
and silent states is observed more often, especially considering the transient firings in the otherwise
silent sleep state. Furthermore, this random firing appears only in the neuron B, but is insufficient to also 
induce spiking in the $\mathrm{A}_1$ neuron. This activity may represents intermittent awakenings, 
which are likely due to the ability of noise to favor the ignition of spiking events.
Such random spikes are not observed when noise acts on the neurons $\mathrm{A}_i$ 
only, even at much larger noise intensities. This may be related to the 
coupling between the neurons $\mathrm{A}_i$, which constrains them
in the same (spiking or silent) state. In order to excite all neurons
$\mathrm{A}_i$ together one would need an input signal affecting all of them in the
same way, which is highly improbable in a realistic system.

The dependence of the coefficient $r$ on $D_\mathrm{B}$ is shown in
Fig.~\ref{fig_rasterB-vs-DB}-B. Again, only a mild stochastic
resonance behavior is suggested  by the data when varying the noise
intensity. 
It should be noted that in this particular configuration
with noise acting only on the neuron B, the response of the neuron B does not
depend on the number $N_\mathrm{A}$ of homogeneous neurons
$\{\mathrm{A}_i\}$, due to the equivalence to the configuration of
the two-neuron model, as we have checked numerically.
Thus, the plots of neuron $\mathrm{A}_1$ in Fig.~\ref{fig_evolution-DB}
are representative of all other neurons $\mathrm{A}_i$. 
In fact, the external current $I_\mathrm{ext}(t)$
as well as the coupling currents are the same for
each neuron $\mathrm{A}_i$, which produces the same response.
According to the equations of the heterogeneous
model, the effective current acting on the neuron B is the arithmetic average of the currents
coming from the various neurons $\{\mathrm{A}_i\}$ and, therefore, coincides with that
of any single neuron $\mathrm{A}_i$.
We use here a homogeneous multi-neuron generalized model only for a
better comparison and consistency with the
rest of this study.

\begin{figure}[!ht]
\begin{flushleft}
\includegraphics[width=8cm]{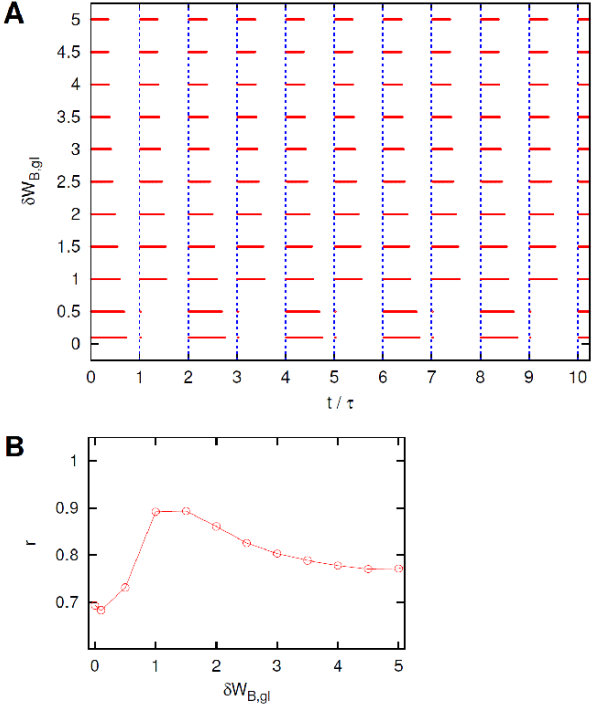}
\end{flushleft}
 \caption{
 {\bf Effect of diversity in the $\mathrm{B}\!\!\to\!\!\mathrm{A}_i$ synapses}.
 (A). Sample of ten periods of the raster plots of neuron B for different
 heterogeneity levels $\delta W_\mathrm{B,gl}$ in the  $\mathrm{B}\!\!\to\!\!\mathrm{A}_i$
 glutamate synapse thresholds, see text for details.
 (B).  Quality of the sleep-wake cycle from the coefficient $r$, Eq.~(\ref{r2}),
 for various threshold diversities $\delta W_\mathrm{B,gl}$.
 \label{fig_rasterB-vs-dWB}
 }
\end{figure}

\subsection{Effects of heterogeneity}
\label{heterogeneity}

The effects introduced by a heterogeneity in the
neurons are dramatic compared to the effects of noise. 
The corresponding improvement of the system response for suitable
intermediate amounts of diversity can be detected very clearly.
This is the main result of this paper and it is
illustrated in this section. Noiseless neurons are assumed 
for easier estimation of the heterogeneity effects  
($D_\mathrm{A} \!=\! D_\mathrm{B} \!=\! 0$).

As in the study of noise described above, we carry out the study of diversity
starting from the same configuration with a non-optimal
double-periodic response to the external periodic stimulus,
corresponding to a zero diversity (homogeneous system).
Heterogeneity is then introduced in the glutamate-induced currents,
either in the thresholds $W^{\,(i)}_{\mathrm{A,gl}}$ regulating the
response of the $\mathrm{A}_i\!\!\to\!\!\mathrm{B}$ synapses at the neuron B 
or in the thresholds $W^{\,(i)}_{\mathrm{B,gl}}$ of the
$\mathrm{B}\!\!\to\!\!\mathrm{A}_i$ synapses at the neurons A.
This is done by randomly extracting values $W$ from a probability
density $f_\mathrm{p}(W)$ and assigning them to the threshold
parameters $W^{\,(i)}_{\mathrm{p,gl}}$ 
($p = \mathrm{A}, \mathrm{B}$).
The probability density used here has a bell-shape
$f_\mathrm{p}(W) = P((W - \overline{W}_\mathrm{p,gl})/\delta W_\mathrm{p,gl})$,
where $P(x) \propto 1/\cosh(x)^2$,
the quantity $\overline{W}_\mathrm{p,gl} = \langle W \rangle$ represents the
average value,
while $\delta W_\mathrm{p,gl}$ measures the dispersion
of the distribution $f_\mathrm{p}(W)$ 
around the average value and is related to the standard deviation
$\sigma_\mathrm{p}$  by 
$\delta W_\mathrm{p,gl} = \pi\sigma_\mathrm{p}/\sqrt{12}$.
For further details see the supporting information in the Appendix.
The width $\delta W_\mathrm{p,gl}$ is assumed 
in the following as the measure of neuronal diversity.
In order to carry out meaningful comparisons with the homogeneous
(two-neuron) model, the average values are set equal to the
corresponding parameters of the homogeneous two-neuron model,
\begin{equation}
  \overline{W}_\mathrm{p,gl} \equiv \int dW \, W \, f_p(W) = W_{\mathrm{p,gl}} \, ,  
                     ~~~~~~p = \mathrm{A, B}.
\end{equation}
The other parameters are unchanged compared to the two-neuron model,
see  Table \ref{table1}.

\begin{figure*}[!ht]
\begin{center}
\includegraphics[width=7in]{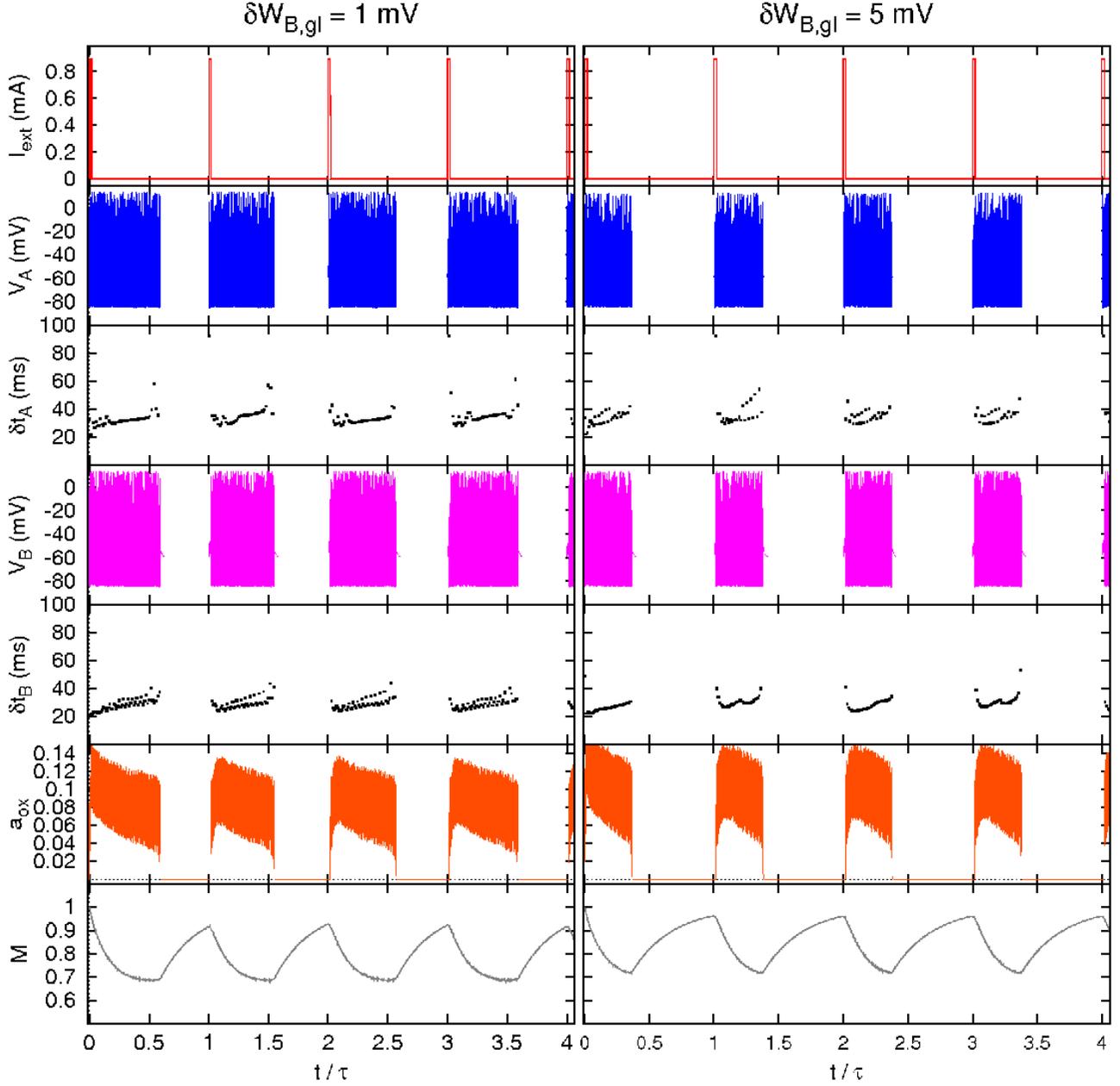}
\end{center}
 \caption{
  {\bf Effect of diversity in the $\mathrm{B}\!\!\to\!\!\mathrm{A}_i$ synapses}.
  Sample of four periods of some relevant variables and inter-spike
  times $\delta t_\mathrm{p}$ ($p = \mathrm{A}, \mathrm{B}$) \textit{versus} time
  for neuron $\mathrm{A}_1$ and neuron B for different levels of the threshold
  diversity $\delta W_\mathrm{B,gl} = 1~\mathrm{mV}$ (left) and $\delta
  W_\mathrm{B,gl} = 5~\mathrm{mV}$ (right).
  Compare Fig.~\ref{fig_rasterB-vs-dWB} and see text for details.
 \label{fig_evolution-dWB}
 }
\end{figure*}

\subsubsection{Diversity in the $\mathrm{B}\!\!\to\!\!\mathrm{A}_i$ synapses (neurons $\mathrm{A}_i$)}

Diversifying the potential thresholds
$W^{\,(i)}_\mathrm{B,gl}$ 
implies heterogeneous glutamate synapses located at the neurons $\mathrm{A}_i$, 
see Eq.~(\ref{agl-N}) and Fig.~\ref{fig_N-neurons-2}.
That is, each neuron $\mathrm{A}_i$ responds in a different way to the
stimulation from the neuron B. 
Notice that this is a truly heterogeneous system which cannot be
reduced to an effective two-neuron model---as in the case of
heterogeneous synapses at neuron B considered in the next section.
We studied a system with $N_\mathrm{A}=20$ neurons
$\mathrm{A}_i$ with diversified threshold parameters
$W^{\,(i)}_\mathrm{B,gl}$, $i = 1, \dots, N_\mathrm{A}$.
The system dynamics were examined for different sets of thresholds
$\{W^{\,(i)}_\mathrm{B,gl}\}$ extracted from distributions $f_\mathrm{B}(W)$  
with different widths $\delta W_\mathrm{B,gl}$ but always the
same average value $\overline{W}_\mathrm{B,gl} = W_\mathrm{B,gl}$.

\begin{figure*}[!ht]
\begin{center}
\includegraphics[width=7in]{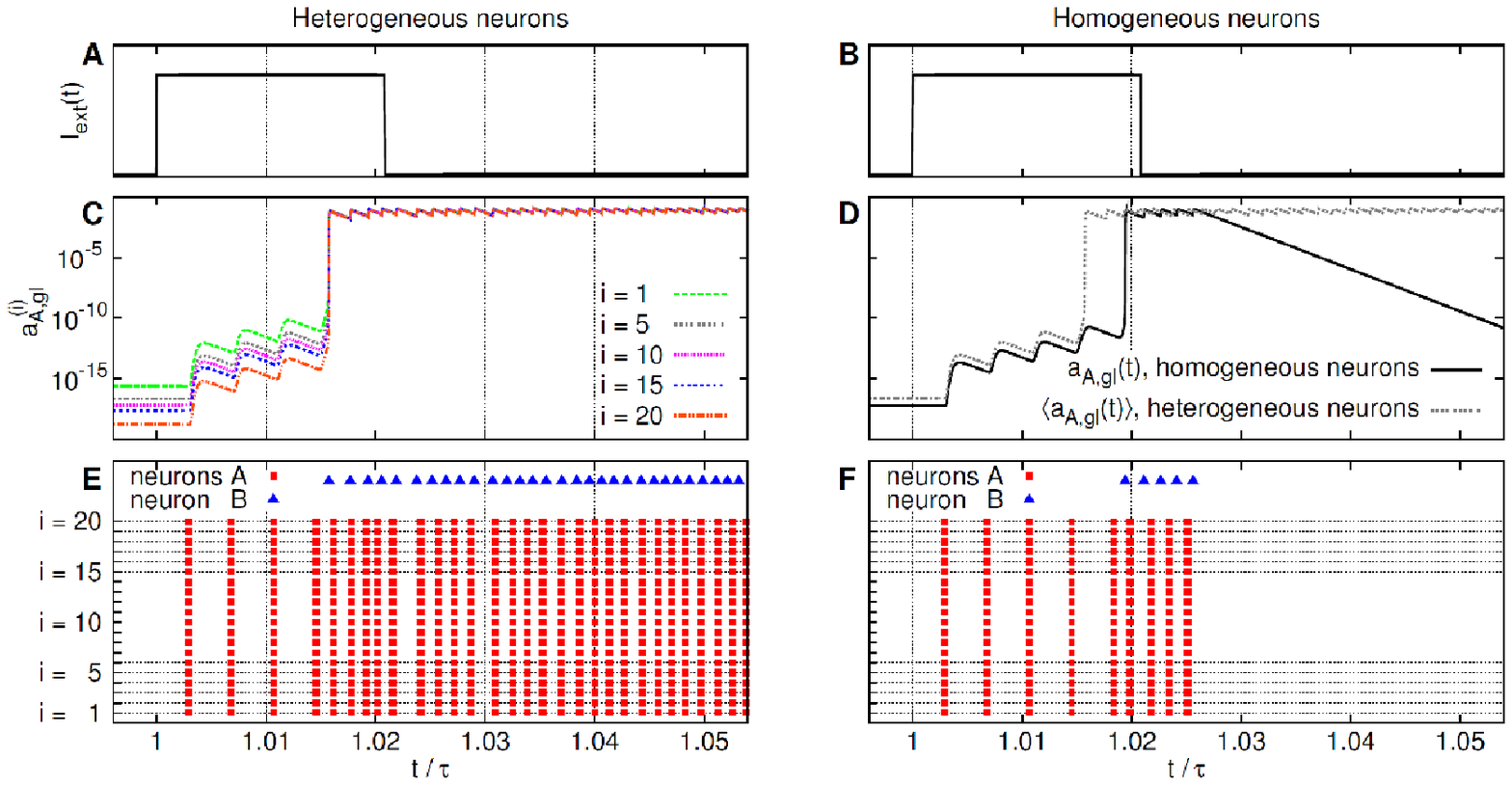}
\end{center}
\caption{
  {\bf Effect of diversity in the $\mathrm{B}\!\!\to\!\!\mathrm{A}_i$ synapses}. 
  Comparison between the responses of the heterogeneous system (left column) 
  and homogeneous system (right column)
  in the first part of the time period $t/\tau \in (1,2)$ 
  during the action of the $500~\mathrm{ms}$ long current pulse 
  starting at $t_1/\tau = 1$ and ending at $t_2/\tau \approx 1.021$.
  (A) and (B) (top panels). External current pulse.
  (C) and (D) (central panels). 
  Behavior of some representative glutamate
  activation variables of the heterogeneous system, 
  $a^{\,(i)}_\mathrm{A,gl}(t)$ for $i = 1,5,10,15,20$ (panel (C)),
  and the (common) time dependence $a_\mathrm{A,gl}(t)$ of the
  homogeneous system activation variables (panel (D), black continuous
  curve); in the latter figure also the average value 
  $\langle a_\mathrm{A,gl}(t) \rangle = N_\mathrm{A}^{-1} \sum_i a^{\,(i)}_\mathrm{A,gl}(t)$ 
  of the heterogeneous system (dashed grey curve) is shown for comparison.
  (E) and (F) (bottom panels). Raster plots of all the neurons of the system.
  See text for further details.
 \label{fig_compare-dWB}
 }
\end{figure*}

The resulting raster plots of the activity of the neuron B are shown in
Fig.~\ref{fig_rasterB-vs-dWB}-A, and a sample of time dependencies 
for the neurons $\mathrm{A}_1$ and B is shown in
Fig.~\ref{fig_evolution-dWB}.
The existence of an optimal degree of diversity, 
corresponding to a value $\delta W_\mathrm{B,gl}$ approximately
between $1$ and $1.5 \mathrm{mV}$, can be clearly seen both from
Fig.~\ref{fig_rasterB-vs-dWB}-A and from the dependence of the
coefficient $r$ on the diversity degree $\delta
W_\mathrm{B,gl}$, in Fig.~\ref{fig_rasterB-vs-dWB}-B.

The underlying mechanism leading the system from the double- to
the single-periodic response as diversity is increased 
can be interpreted following the prototype mechanical model of
diversity-induced resonance introduced in Ref. \cite{Tessone2006a}.
In this model a set of interacting oscillators moving in a
bistable potential is subjected to an external periodic force,
which pushes the system toward the left and the right barrier
alternately.
If the oscillators are identical, i.e. they have the same parameter values
corresponding to an under-threshold regime, 
then the system of oscillators cannot perform jumps on the
other site of the  barrier under the action of the applied periodic force.
However, when the parameters are diversified (keeping constant the
corresponding average value) some oscillators respond more promptly
to the force and jump to the other side of the barrier, gradually
pulling the rest of the system.
In the present case, each neuron $\mathrm{A}_i$ corresponds to a
nonlinear oscillator of the example, while the parameter which is
diversified is the activation thresholds $W^{\,(i)}_{\mathrm{B,gl}}$
of the glutamate-induced currents.

\begin{figure}[!ht]
\begin{flushleft}
\includegraphics[width=8cm]{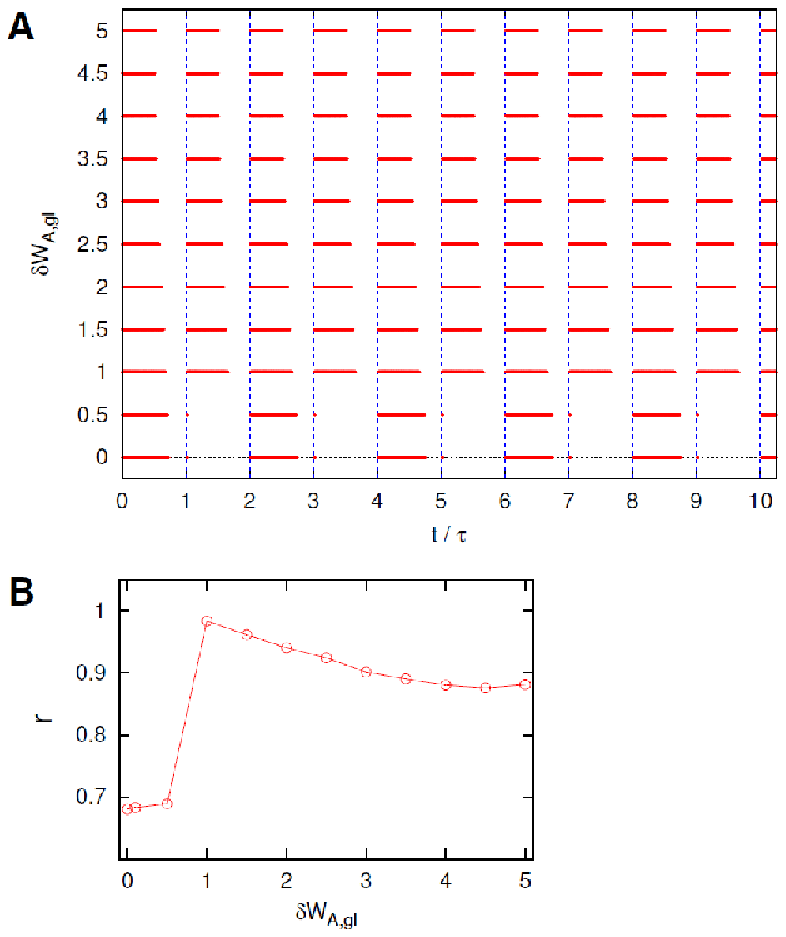}
\end{flushleft}
 \caption{
 {\bf Effect of diversity in the $\mathrm{A}_i\!\!\to\!\!\mathrm{B}$ synapses.}
 (A). Sample of ten periods of raster plots of neuron B for different
 heterogeneity levels $\delta W_\mathrm{A,gl}$ in the $\mathrm{A}_i\!\!\to\!\!\mathrm{B}$
 glutamate synapse thresholds, see text for details.
 (B). Coefficient $r$, Eq.~(\ref{r2}), for various degrees of diversity $\delta W_\mathrm{A,gl}$.
 \label{fig_rasterB-vs-dWA}
 }
\end{figure}

To show that this is the actual mechanism in action,
Fig.~\ref{fig_compare-dWB} (left) illustrates the response of the
heterogeneous system by depicting the time dependence of the glutamate
activation variables $a^{\,(i)}_\mathrm{A,gl}(t)$ of the neurons
$\mathrm{A}_i$, $i = 1, 5, 10, 15, 20$, with
different values of the thresholds $W^{\,(i)}_{\mathrm{B,gl}}$, at the
beginning of a new period in the presence of the periodic current pulse.
In Fig.~\ref{fig_compare-dWB} also the raster plots for all
neurons in the same time interval are shown.
One can notice that the activation variables
$a^{\,(i)}_\mathrm{A,gl}(t)$ behave differently from each other. 
Those associated to the lowest values of the activation threshold (indicated by small $i$ values) respond
stronger to the current pulse than those with the highest values of the
threshold (largest values of $i$).
The system is observed to reach the spiking regime faster
than in the homogeneous case, which is shown in the right part of
Fig.~\ref{fig_compare-dWB} through the comparison between the
glutamate activation variable of the homogeneous system,
$a^{\,}_\mathrm{A,gl}(t)$, and the average activation variable 
$\langle a_\mathrm{A,gl}(t) \rangle = N_\mathrm{A}^{-1} \sum_i a^{\,(i)}_\mathrm{A,gl}(t)$ 
of the heterogeneous system.
Eventually, $a^{\,}_\mathrm{A,gl}(t) \to 0$ and the homogeneous system
goes back to the silent state, while the average activation variables
of the heterogeneous system 
(and their average $\langle a^{\,(i)}_\mathrm{A,gl}(t) \rangle$)
continue to oscillate around positive values, signaling the
stability of the reached firing state.

\begin{figure*}[!ht]
\begin{center}
\includegraphics[width=7in]{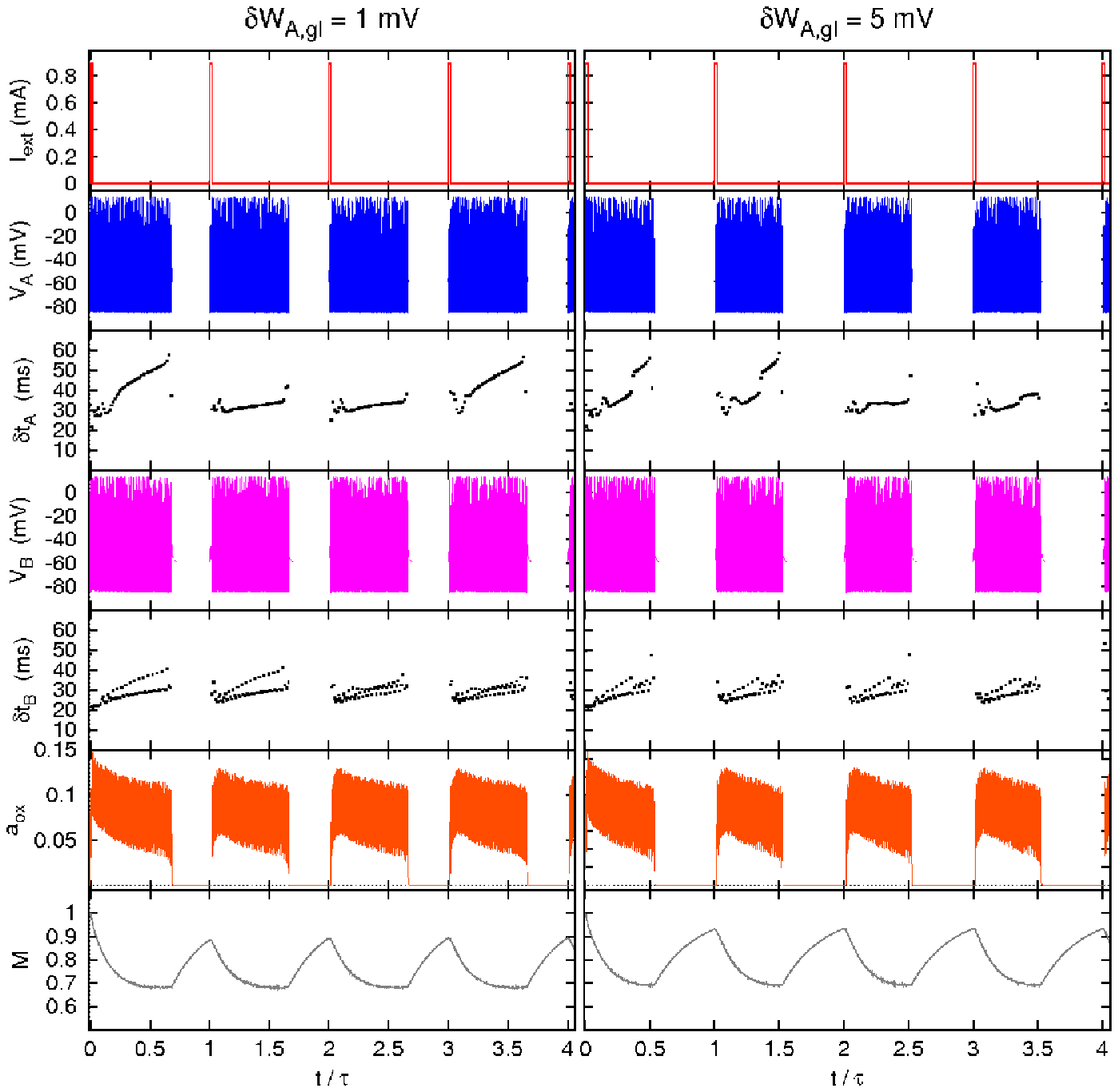}
\end{center}
 \caption{
  {\bf Effect of diversity in the $\mathrm{A}_i\!\!\to\!\!\mathrm{B}$}
  synapses.
  Sample of four periods of the time dependence of some relevant system
  variables and inter-spike times $\delta t_\mathrm{p}$
  ($p = \mathrm{A}, \mathrm{B}$) of neuron $\mathrm{A}_1$ and neuron B
  for a threshold diversity $\delta W_\mathrm{A,gl} = 1~\mathrm{mV}$ (left) and
  $W_\mathrm{A,gl} = 5~\mathrm{mV}$ (right).
  Compare Fig.~\ref{fig_rasterB-vs-dWA} and see text for details.
 \label{fig_evolution-dWA}
 }
\end{figure*}

\subsubsection{Diversity in the $\mathrm{A}_i\!\!\to\!\!\mathrm{B}$ synapses (neuron \textrm{B})}

In order to study the effects of added heterogeneity in the glutamate
synapses located at the neuron B, one has to diversify the potential
threshold parameters $W^{\,(i)}_\mathrm{A,gl}$, see Eq.~(\ref{bgl-N}).
For this particular case, it is possible to simplify the model into a
two-neuron model with a single effective AB coupling.
This is possible because 
heterogeneity only enters Eq.~(\ref{bgl-N}), while
other model equations reduce to the same equations in the case
of identical neurons $\mathrm{A}_i$, so that all neurons
$\mathrm{A}_i$ behave in the same way.
Thus, the effective glutamate-induced current to the neuron B is 
\begin{eqnarray}
  I_\mathrm{eff} (V)
  &=& 
  \frac{1}{N_\mathrm{A}} \sum_{i=1}^{N_\mathrm{A}} 
  \left[
    \Phi(S_\mathrm{gl} (V_\mathrm{A} \!-\! W^{\,(i)}_\mathrm{A,gl}))
  \right]
  \nonumber \\
  &\approx& 
  \int dW \, f(W) \, \Phi(S_\mathrm{gl} (V_\mathrm{A} \!-\! W))
  \, ,
  \label{Ieff}
  \end{eqnarray}
where $V_\mathrm{A}$ is the common value of the membrane potentials of the
neurons $\mathrm{A}_i$.
Here $f(W)$ is the probability density of the
corresponding thresholds $W = W^{\,(i)}_\mathrm{A,gl}$,
which is assumed to be the same bell-shaped probability density 
$f(W) = P((W - \overline{W})/\delta W)$ as discussed above, with
$\overline{W} = W^{\,}_\mathrm{A,gl}$ and
$\delta W = \delta W^{\,}_\mathrm{A,gl}$.

We now consider two limiting cases of the effective current given by
Eq.~(\ref{Ieff}).
In the limit $\delta W \ll S^{-1}_\mathrm{gl}$,
when diversity is very small on the scale $S^{-1}_\mathrm{gl}$, 
it can be assumed that the following approximation holds, 
$f(W) = P((W - \overline{W})/\delta W) \approx \delta(W - \overline{W})$ and the
integral (\ref{Ieff}) can be reduced to the homogeneous result,
\begin{eqnarray}
  I_\mathrm{eff} (V)
  &\approx& 
  \int dW \, \delta(W - \overline{W}) \, \Phi(S_\mathrm{gl} (V_\mathrm{A} \!-\! W))
  \nonumber \\
  &=& \Phi(S_\mathrm{gl} (V_\mathrm{A} \!-\! \overline{W})) 
  \, .
  \label{Ieff2}
  \end{eqnarray}
In the complementary limit of high diversity level, 
$\delta W \gg S^{-1}_\mathrm{gl}$, 
the smooth function 
$\Phi(S_\mathrm{gl} (V_\mathrm{A} \!-\! W))$
can be approximated with Heaviside step functions 
$\Theta(W \!-\! V_\mathrm{A})$,
and the effective current becomes
\begin{eqnarray}
  I_\mathrm{eff} (V)
  &\approx& 
  \int dW \, f(W) \, \Theta(W \!-\! V_\mathrm{A})
  \nonumber \\
  &=& \int dW \, P((W - \overline{W})/\delta W) \, \Theta(W \!-\! V_\mathrm{A})
  \nonumber \\
  &=& \Phi((V_\mathrm{A} \!-\! \overline{W})/\delta W)
  \, .
  \label{Ieff3}
  \end{eqnarray}
This follows from the form of the chosen distribution, 
$P(x) \propto 1/\cosh(x)^2 = - 4\,d\Phi(x)/dx$.
Thus, in both these limiting cases the effective current can be written in the form  
$I_\mathrm{eff} (V) \approx \Phi(\lambda(V_\mathrm{A} \!-\! \overline{W}))$, 
with $\lambda = S_\mathrm{gl}$ 
for $\delta W \ll S^{-1}_\mathrm{gl}$ and 
$\lambda = \delta W^{-1}$ 
for $\delta W \gg S^{-1}_\mathrm{gl}$.
It is interesting that, 
as we have checked by numerical integration of Eq.~(\ref{Ieff}),
the same analytical form also holds for intermediate values of $\delta
W$ and $S^{-1}_\mathrm{gl}$, so to a very good approximation the
effective current can be written as
$I_\mathrm{eff} (V) \approx \Phi(\lambda(V_\mathrm{A} \!-\! \overline{W}))$,
where the parameter $\lambda$ depends on the ratio $\delta
W/S^{-1}_\mathrm{gl}$ and varies monotonously between
$S_\mathrm{gl}$ and $\delta W^{-1}$, as $\delta W/S^{-1}_\mathrm{gl}$ varies between 0 and $\infty$.

The system's response at different levels of heterogeneity,
$\delta W_\mathrm{A,gl}$, is presented in
Fig.~\ref{fig_rasterB-vs-dWA}-A through the raster plots for the neuron B.
A sample of time dependence is shown in
Fig.~\ref{fig_evolution-dWA}, while Fig.~\ref{fig_rasterB-vs-dWA}-B shows the
dependence of the coefficient $r$ on the diversity level.
In Fig.~\ref{fig_rasterB-vs-dWA}, it can be seen that 
for small values of the diversity $\delta W_\mathrm{A,gl}$
the response of the neuron B presents the double periodicity of the
reference configuration.
Single periodicity is recovered for higher levels of diversity.
At even higher values of $\delta W_\mathrm{A,gl}$, 
the coefficient $r$ begins to decrease.
The points of the curve corresponding to the highest values of
$r$ suggest an optimal degree of diversity 
$\delta W_\mathrm{A,gl} \approx 1~\mathrm{mV}$.

The main difference compared to the case in which noise
intensity is varied, is that the response of the system remains more
regular also at the highest levels of diversity considered,
i.e. without random spikes appearing during the silent state and with
a typical cycle well shared between a day and a night sub-period.

\section{Discussion}

In the present work we have introduced a heterogeneous multi-neuron version of 
the previously developed physiologically motivated model of the homeostatic regulation of sleep. 
The multi-neuron model is composed of a population of conductance-based orexin-producing neurons
and a single representative glutamatergic neuron. In this model the glutamatergic and orexinergic 
neurons are undergoing transitions between firing and silence depending on the external circadian input 
and internal homeostatic mechanisms. These transitions correspond to the transitions between wake (firing) 
and sleep (silence), with the homeostatic mechanism being dependent on the availability of orexin. 

The specific aim of this study was to explore the effects of noise and diversity 
in the regulation of sleep-wake cycles in such a model. It is clear that diversity and noise  
are integral parts of all biological systems, including the orexinergic neuronal population 
in the lateral hypothalamus. However, the role of disorder, and especially diversity, 
is rarely considered in the physiologically based mathematical models of sleep-related systems 
\cite{Bazhenov2002a,Hill2005a,DinizBehn2007a,Best2007a,DinizBehn2008a,Booth2011a,Williams2011a}. 
To our knowledge, diversity had so far been included 
only in one such model, i.e. the model of interacting 
circadian oscillators  \cite{Komin2010a}, and here we present another example of the 
constructive role of diversity in regulation of sleep. 

We have demonstrated the existence of a diversity-induced
resonance, leading to a clear and strong improvement of the quality
of the sleep-wake cycles, at a physiologically justified intermediate
level of diversity of the orexin-producing neurons. However, only a mild improvement was found 
with varying noise intensity (stochastic resonance phenomena). 

We have considered the simplest system with only 20 heterogeneous orexin neurons 
and one local glutamate neuron. Also we have used a very simple all-to-all network topology for the 
connections among orexinergic neurons. However, it can be expected that constructive 
effects of diversity will be found also in other model configurations. 
In the future, more realistic modifications of the model with a larger population 
of glutamatergic neurons and more sophisticated inter-populations 
connections should be considered. Furthermore, in the future studies interplay 
between noise and diversity should likewise be investigated, since in 
nature both types of disorder are normally present.  

The validity of the result obtained within this model may be more
general, since diversity-induced resonance is known to take place for
suitable values of the parameters in general networks of interacting
(non-linear) oscillators. A question then naturally arises: 
whether the phenomena encountered here
could also characterize other systems where there is a coupling
between two very different time scales or, in other words, if
homeostatically regulated biological systems may take advantage from a
suitable level of heterogeneity of their components.

\newpage

\begin{acknowledgments}
We acknowledge financial support from the EU NoE BioSim,
LSHB-CT-2004-005137, and project FIS2007-60327 (FISICOS) from MINECO
and FEDER (Spain). M.P. acknowledges financial support from the
Estonian Ministry of Education and Research through Project
No. SF0690030s09 and the Estonian Science Foundation via grants
no. 7466 and no. 9462. S.P. acknowledges funding from ARC and NHMRC.
\end{acknowledgments}

\appendix

\section{Supporting information: further details on the numerical simulation}

\subsection{Evaluation of the wake and sleep times}
\label{sleeptimes}
The quality of a sleep-wake cycle was estimates through the coefficient
$r = \Delta t^{(1)}/\tau_1 - \Delta t^{(2)}/\tau_2$,  
where $\Delta t^{(\alpha)}$
represents the wakefulness time interval during the day ($\alpha = 1$) 
or night ($\alpha = 2$),
while $\tau_1$ and $\tau_2$ represent the length of day and night,
respectively.
This coefficient can vary between a minimum value $r = -1$, representing an
exchange of wakefulness and sleep between day and night, and a maximum
value $r = 1$, corresponding to the optimal situation with wakefulness during
the whole day time interval $\Delta t^{(1)} = \tau_1$ and
continuous sleep during the whole night interval, $\Delta t^{(2)} = 0$.
For a more realistic estimate of the quality of the sleep-wake cycle
the two wakefulness time intervals $\Delta t^{(1)}$ and $\Delta t^{(2)}$ 
were  computed in slightly different ways, as described below.

\vspace{0.5cm}

{\bf Estimating wakefulness during the day: $\Delta t^{(1)}$}. 
In order to estimate the wakefulness during the day, 
when an individual is supposed to be awake,
we have considered the quality of wakefulness by computing $\Delta
t^{(1)}$ from states characterized by an ``effective
wakefulness'' in a tonic firing regime.
On the other hand, isolated spikes, which break a sleep period, 
did not contribute to the wakefulness interval and were neglected:
such isolated spikes can at most represent a fragmented
wakefulness state, similar to that of a narcoleptic individual.
A tonic spiking state was recognized
checking if the inter-spike time was smaller than a suitable
threshold $\tau_\mathrm{max}$; in this case such an inter-spike interval
was added to the total wakefulness time interval 
$\Delta t_\mathrm{wake}^{(1)}$.
On the other hand, 
if the corresponding inter-spike time was larger than the threshold
$\tau_\mathrm{max}$, then the two corresponding spikes were considered to be
isolated and that time interval was neglected.
Notice that in general inter-spike times are not universal and vary with external
parameters; for this reason a value $\tau_\mathrm{max} = 100$ ms was chosen
heuristically, considering the typical working conditions of the system.

\vspace{0.5cm}

{\bf Estimating wakefulness during the night: $\Delta t^{(2)}$}. 
As for the night is concerned, it is the quality of sleep which
contributes to the overall quality of the sleep-wake cycle.
Therefore, on the difference of $\Delta t^{(1)}$, 
we have computed $\Delta t^{(2)}$,
the wakefulness period during night,
including also short wakefulness events (isolated spikes), since they break the
sleep period, thus worsening the quality of sleep.
Each isolated spike is assumed to contribute a conventional time
interval, assumed as $\tau_\mathrm{max}$ for simplicity.

The difference between the two mentioned ways to evaluate wakefulness
periods is relevant at high level of noise, which produces many isolated spikes.

\vspace{0.5cm}

\subsection{Extraction of parameters with a given distribution}
\label{extraction}
The different values of the glutamate thresholds $\{W_i\}$ of neurons
$\{\mathrm{A}_i\}$, 
which make neurons A heterogeneous, 
were extracted using the probability distribution
\begin{equation}
  \label{fx}
  f(W) = 
  \frac
  {1} 
  {2 \, \delta W \, \mathrm{cosh}^2[(W - \bar{W})/\delta W]}
  \, .
  \end{equation}
Here $\bar{W} = \int dW \, W\, f(W)$ is the average value and
the parameter $\delta W$ measures the dispersion of the distribution
around $\bar{W}$.
It is proportional to the standard deviation $\sigma$,
namely, it is related to the variance according to 
$\sigma^2 \equiv \langle (W - \bar{W})^2 \rangle = \pi^2 \delta W^2 / 12$.
The function (\ref{fx}) can be integrated to obtain the lower
cumulative distribution function $F(x)$,
\begin{eqnarray}
  \label{Fx}
  F(W) 
  &=& \frac{1}{2} \left\{ 1 + \tanh[(W - \bar{W})/\delta W] \right\}
  \nonumber \\
  &=& \frac{1}{1^{\,} + \exp[-2(W - \bar{W})/\delta W]}
  \, .
  \end{eqnarray}
Simulations were made by using various sets of A-neuron thresholds $\{W_i\}$
obtained by rescaling the values of a same set of thresholds by $\delta W$. 
This can be done for example by inverting the same uniform distribution of values
$\{F_i\}$ and using distributions $F(W)$ corresponding to the
different desired values of $\delta W$. 
For all the parameter sets  $\{W_i\}$ used the average value $\langle
W \rangle$ was the same. 
For a few set of parameters we have checked that equivalent results
are obtained by first extracting randomly the parameters from the
distribution $F(x)$ with the desired $\delta W$ and then averaging the
final results obtained.
In the latter case the set of values $\{F_i\}$ are extracted randomly
in the interval $F_i \in (0,1)$.

\newpage
\newpage


\end{document}